\documentclass[10pt,journal, letterpaper]{IEEEtran}
\usepackage[table]{xcolor}
\usepackage{algorithm,algorithmic,setspace,amsfonts,amsmath,amssymb,bbm,color,comment,epsfig,float,graphics,enumitem,epstopdf,amsthm,mathtools,multirow,multicol,subcaption,tikz,url}
\usepackage[sort,noadjust]{cite}
\usepackage[acronym]{glossaries}
\usepackage{hyperref}
\usepackage{makecell}
\usepackage[figurename=Fig.,font=small,labelsep=period]{caption}

{\newtheorem{theorem}{Theorem}}
{}
{}
{\newtheorem{proposition}[theorem]{Proposition}}
{}
{}
{}

\newcommand{\lb}{\rm{lb}}

\newcommand{\name}{covariance shaping}
\newcommand{\Name}{Covariance shaping}

\newcommand{\rmw}{\textrm{w}}
\newacronym[plural=UEs, firstplural=user equipments (UEs)]{ue}{UE}{user equipment}
\newacronym{bs}{BS}{base station}
\newacronym{bd}{BD}{block diagonalization}
\newacronym{d2d}{D2D}{device-to-device}
\newacronym{csit}{CSIT}{channel state information at the transmitter}
\newacronym{csi}{CSI}{channel state information}
\newacronym{los}{LoS}{line-of-sight}
\newacronym{nlos}{NLoS}{non-line-of-sight}
\newacronym{ula}{ULA}{uniform linear array}
\newacronym{upa}{UPA}{uniform planar array}
\newacronym{mimo}{MIMO}{multiple-input multiple-output}
\newacronym{miso}{MISO}{multiple-input single-output}
\newacronym{mmse}{MMSE}{minimum mean squared error}
\newacronym{mrt}{MRT}{maximum ratio transmission}
\newacronym{nr}{NR}{New Radio}
\newacronym[plural=AoAs, firstplural=angles of arrival (AoAs)]{aoa}{AoA}{angle of arrival}
\newacronym{slnr}{SLNR}{signal-to-leakage-and-noise ratio}
\newacronym{sinr}{SINR}{signal-to-interference-plus-noise ratio}
\newacronym{snr}{SNR}{signal-to-noise ratio}
\newacronym{se}{SE}{spectral efficiency}
\newacronym{fdd}{FDD}{frequency-division duplex}
\newacronym{tdd}{TDD}{time-division duplex}
\newacronym{5g}{5G}{fifth-generation}
\newacronym{mse}{MSE}{mean squared error}
\newacronym{zf}{ZF}{zero-forcing}
\newacronym{wf}{WF}{water-filling}
\newacronym{nmse}{NMSE}{normalized MSE}
\newacronym{rzf}{RZF}{regularized zero-forcing}
\newcommand{\rmBS}{\textnormal{\tiny{BS}}}
\newcommand{\rmUE}{\textnormal{\tiny{UE}}}
\DeclarePairedDelimiter{\ceil}{\lceil}{\rceil}
\newcommand{\rmm}[1]{\mathrm{#1}}

\makeatletter
\let\oldabs\abs
\def\abs{\@ifstar{\oldabs}{\oldabs*}}

\let\oldnorm\norm
\def\norm{\@ifstar{\oldnorm}{\oldnorm*}}
\makeatother
\renewcommand{\a}{\mathbf{a}}
\renewcommand{\b}{\mathbf{b}}

\newcommand{\e}{\mathbf{e}}

\newcommand{\g}{\mathbf{g}}
\newcommand{\h}{\mathbf{h}}

\newcommand{\p}{\mathbf{p}}

\newcommand{\s}{\mathbf{s}}

\renewcommand{\u}{\mathbf{u}}
\renewcommand{\v}{\mathbf{v}}
\newcommand{\w}{\mathbf{w}}
\newcommand{\x}{\mathbf{x}}
\newcommand{\y}{\mathbf{y}}
\newcommand{\z}{\mathbf{z}}

\newcommand{\0}{\mathbf{0}}

\newcommand{\A}{\mathbf{A}}

\renewcommand{\H}{\mathbf{H}}
\newcommand{\I}{\mathbf{I}}

\renewcommand{\P}{\mathbf{P}}
\newcommand{\Q}{\mathbf{Q}}
\newcommand{\R}{\mathbf{R}}

\newcommand{\T}{\mathbf{T}}
\newcommand{\U}{\mathbf{U}}
\newcommand{\V}{\mathbf{V}}
\newcommand{\W}{\mathbf{W}}

\newcommand{\Y}{\mathbf{Y}}
\newcommand{\Z}{\mathbf{Z}}

\newcommand{\Lambdab}{\mathbf{\Lambda}}

\newcommand{\Sigmab}{\mathbf{\Sigma}}

\newcommand{\Phib}{\mathbf{\Phi}}

\newcommand{\setC}{\mathcal{C}}

\newcommand{\setN}{\mathcal{N}}
\newcommand{\setO}{\mathcal{O}}

\newcommand{\setS}{\mathcal{S}}

\newcommand{\Real}{\mbox{$\mathbb{R}$}}
\newcommand{\Compl}{\mbox{$\mathbb{C}$}}

\newcommand{\argmin}{\operatornamewithlimits{argmin}}

\newcommand{\Exp}{\mathbb{E}}
\newcommand{\rmF}{\mathrm{F}}
\newcommand{\herm}{\mathrm{H}}

\newcommand{\minimize}{\operatornamewithlimits{minimize}}

\newcommand{\subjectto}{\mathrm{subject~to}}
\newcommand{\tr}{\mathrm{tr}}
\newcommand{\tran}{\mathrm{T}}
\newcommand{\Var}{\mathbb{V}}

\title{Enforcing Statistical Orthogonality in Massive MIMO Systems via Covariance Shaping}
\author{Placido~Mursia,~\IEEEmembership{Member,~IEEE,} Italo~Atzeni,~\IEEEmembership{Member,~IEEE,} Laura~Cottatellucci,~\IEEEmembership{Member,~IEEE,} \\ and David~Gesbert,~\IEEEmembership{Fellow,~IEEE}

\thanks{P.~Mursia is with NEC Laboratories Europe GmbH, Heidelberg, Germany, and the Communication Systems Department, EURECOM, France (email: placido.mursia@neclab.eu). I.~Atzeni is with the Centre for Wireless Communications, University of Oulu, Finland (email: italo.atzeni@oulu.fi). L.~Cottatellucci is with the Department of Electrical, Electronics, and Communication Engineering, Friedrich-Alexander University Erlangen-Nuremberg, Germany (email: laura.cottatellucci@fau.de). D.~Gesbert is with the Communication Systems Department, EURECOM, France (email: david.gesbert@eurecom.fr).}

\thanks{The work of P.~Mursia was supported by the Marie Sk\l{}odowska-Curie Actions (MSCA-ITN 722788 SPOTLIGHT). The work of I.~Atzeni was supported by the Marie Sk\l{}odowska-Curie Actions (MSCA-IF 897938 DELIGHT). Part of this work has been presented at IEEE GLOBECOM 2018~\cite{Mur18} and at IEEE CAMAD 2018~\cite{Mur18_2}.}
}

\begin{document}

\maketitle

\begin{abstract}
This paper tackles the problem of downlink data transmission in massive multiple-input multiple-output (MIMO) systems where  user equipments (UEs) exhibit high spatial correlation and channel estimation is limited by strong pilot contamination. Signal subspace separation among  UEs is, in fact, rarely realized in practice and is generally beyond the control of the network designer (as it is dictated by the physical scattering environment). In this context, we propose a novel statistical beamforming technique, referred to as \textit{MIMO covariance shaping}, that exploits multiple antennas at the UEs and leverages the realistic non-Kronecker structure of massive MIMO channels to target a suitable shaping of the channel statistics performed at the UE-side. To optimize the covariance shaping strategies, we propose a low-complexity block coordinate descent algorithm that is proved to converge to a limit point of the original nonconvex problem. For the two-UE case, this is shown to converge to a stationary point of the original problem. Numerical results illustrate the sum-rate performance gains of the proposed method with respect to spatial multiplexing in scenarios where the spatial selectivity of the base station is not sufficient to separate closely spaced UEs.

\textbf{\textit{Index terms}---Covariance shaping, massive MIMO, multi-user MIMO, pilot contamination, statistical beamforming.}
\end{abstract}

\glsresetall

\section{Introduction}

Massive \gls{mimo} is a multi-antenna technology that has great potential to boost the \gls{se} of cellular networks by means of highly directional beamforming and spatial multiplexing of many \glspl{ue} in the same time-frequency resources. It thus plays a pivotal role in current 5G \gls{nr} implementations \cite{Boc14,Lar14,And14} and is expected to maintain this prominence in future wireless generations \cite{Raj20,Zha20,Che20,Ngo20,Atz21}. The benefits of massive \gls{mimo} can be ascribed to the large number of antennas available at the \gls{bs}, which we denote by $M$. In this context, it is shown in \cite{Bjo18} that the achievable \gls{se} of downlink/uplink massive \gls{mimo} systems is unbounded as $M$ grows large and when \gls{mmse} precoding/combining is adopted at the \gls{bs}, which can asymptotically remove any interference. However, in presence of a large number of \glspl{ue} and finite \gls{bs} antennas, the aforementioned approach might still be limited by interference and must rely on accurate instantaneous \gls{csi}. In crowded scenarios, such as outdoor events, transport hubs, and stadiums, the channels of closely spaced \glspl{ue} exhibit high spatial correlation, which hinders the capability of the \gls{bs} to separate such \glspl{ue} during both the channel estimation phase and the data transmission phase. Moreover, when the channel coherence time is limited, non-orthogonal pilots might be used across the \glspl{ue} during the channel estimation phase, which results in strong pilot contamination \cite{Jos11,Mar16}.

To overcome these issues and facilitate the operations in the massive \gls{mimo} regime, several works have proposed to leverage statistical \gls{csi} \cite{Nam12,Adh13,Nam14,Yin13,Yin14,Zha08,Dai15,Liu13,Pad19,You15,Mog17,Yin16}. This consists mainly in the channel covariance matrix of each \gls{ue}, which is essentially dictated by the angle spread spanned by the multipath propagation of the signals impinging on the antenna array. This angle spread is often bounded due to the high spatial resolution of the massive array compared with the limited scattering. As a result, the channel covariance matrices in massive \gls{mimo} tend to be low-rank and dominated by few major propagation directions \cite{Bjo18,Yin13,Yin14}. This particular property can be exploited for several applications such as reducing the feedback overhead in the channel estimation phase \cite{Nam12,Adh13, Nam14} and mitigating the interference in the downlink data transmission phase when the \glspl{ue} exhibit non-overlapping or orthogonal signal subspaces \cite{Yin13,Yin14}. For example, statistical \gls{csi} can be used to precode signals such that their chosen propagation paths do not interfere in average \cite{Zha08,Dai15}. In this regard, a robust precoding/decoding design based on the average \gls{mse} matrix is derived in \cite{Zha08} under different \gls{csi} conditions, whereas \cite{Dai15} provides a lower bound on the ergodic sum rate for the two-\gls{ue} setting when the \glspl{ue} are equipped with a single antenna. 

There is also a large body of literature focusing on the hybrid precoding problem, where the beamforming applied at the \gls{bs} is factorized into an inner and an outer precoding matrix \cite{Nam12,Adh13,Nam14,Liu13,Pad19}. Here, the former is based on instantaneous \gls{csi} while the latter depends only on the second-order channel statistics. This results in reduced pilot length required for the instantaneous effective channel estimation and lower computational complexity associated with the inner precoding design. Moreover, in the context of \gls{fdd} systems, such a hybrid precoding method can be used to reduce the size of the resulting effective channels by exploiting the near-orthogonality of the angle spreads of different \gls{ue} groups \cite{Nam12}. In \cite{Adh13}, this approach is shown to incur a vanishing loss compared with the case of full \gls{csi} when the number of \gls{bs} antennas grows large, while the same result is obtained in \cite{Nam14} for the case of finite number of \gls{bs} antennas and suitable \gls{ue} scheduling. In \gls{tdd} systems, existing works have pointed out how the exploitation of statistical \gls{csi} enables efficient pilot reuse across the \glspl{ue} \cite{You15,Mog17,Yin16}. In \cite{You15}, pilot reuse for massive \gls{mimo} transmissions over spatially correlated Rayleigh fading channels is proposed when the angle spreads of the \glspl{ue} with the same pilots are non-overlapping. In \cite{Mog17}, the second-order channel statistics are used to precode the pilots, thus reducing the variance of the channel estimation error by a factor that is proportional to the number of \gls{ue} antennas. A robust channel estimation method dealing with scenarios in which the angle spreads of the desired and interfering channels are overlapping is proposed in \cite{Yin16} and exploits channel separability in the power domain \cite{Mul14}.

Most of the above works rely on specific structures of the channel covariance matrices of the \glspl{ue} in terms of rank and degree of separation among  signal subspaces. Such properties are determined by the locations of the \glspl{ue} combined with the surrounding scattering environments, where both these factors are generally beyond the control of the network designer. Indeed, in many practical scenarios, the \glspl{ue} exhibit high spatial correlation, e.g., when they are not sufficiently far apart. Hence, the condition of non-overlapping or orthogonal signal subspaces is rarely satisfied in practice \cite{Bjo18}. Note that, in this setting, the number of \gls{bs} antennas needed to effectively separate interfering \glspl{ue} with \gls{mmse} precoding/combining increases with their spatial correlation, i.e., as the \glspl{ue} get closer to each other, and may become impractical in crowded scenarios. Moreover, existing works are often based on the assumption of Kronecker channel model owing to its simple analytical formulation. However, such a model has been shown to be an over-simplification of the true nature of \gls{mimo} channels \cite{Rag10}.\footnote{The Kronecker channel model is justified in scenarios where both transmitter and receiver are surrounded by clusters of scatterers that are very far apart, which is hardly realized in practice \cite{Wei06}.}

\subsection{Contributions}

Building on the fact that most future \glspl{ue} are envisioned to be equipped with a small-to-moderate number of antennas, this paper proposes a novel statistical beamforming technique at the \glspl{ue}-side, referred to as \textit{\gls{mimo} \name}.\footnote{Note that the term ``\name'' has been used in the past in entirely different contexts, mainly estimation theory (see, e.g., \cite{Eld03}).} While existing methods assume statistical orthogonality among the \glspl{ue} as a property given by the physical scattering environment and the Kronecker channel model due to its analytical tractability, \gls{mimo} \name{} aims at modifying the channel statistics of the \glspl{ue} with highly overlapping channel covariance matrices in order to enforce signal subspace separation in scenarios where the spatial selectivity of the \gls{bs} is not sufficient to separate such \glspl{ue}. The proposed approach consists in preemptively applying a statistical beamforming at the \gls{ue}-side during both the uplink pilot-aided channel estimation phase and the downlink data transmission phase, and relies uniquely on statistical \gls{csi}. In this context, each \gls{ue} employs its antennas to excite only a subset of all the possible propagation directions towards the \gls{bs} such that the spatial correlation among the interfering \glspl{ue} is minimized while preserving enough useful power for effective data transmission. Hence, \gls{mimo} \name{} is suitable for both pilot decontamination in \gls{tdd} systems and statistical precoding/combining. Remarkably, the proposed method exploits the realistic non-Kronecker channel structure, which allows to suitably alter the channel statistics perceived at the \gls{bs} by designing the transceiver at the \gls{ue}-side. Therefore, it has the unique advantage of turning the generally inconvenient non-Kronecker nature of massive \gls{mimo} channels into a benefit. Numerical results show the sum-rate performance gains with respect to a reference scheme employing  multiple antennas at the \gls{ue} for spatial multiplexing in scenarios where the spatial selectivity of the \gls{bs} is not sufficient to separate \glspl{ue} with highly overlapping channel covariance matrices and the channel estimation is limited by strong pilot contamination.

The \gls{mimo} \name{} framework was initially proposed in our prior works \cite{Mur18,Mur18_2} and recently used for the minimization of the outage probability in \cite{Has20}. This work extends the metric proposed in \cite{Mur18} to all pairs of interfering \glspl{ue} to design a statistical receive beamforming at the \gls{ue} along with a statistical precoding at the \gls{bs}. Differently from \cite{Has20}, we argue that \gls{mimo} \name{} must be adopted for \emph{both} pilot decontamination during the uplink pilot-aided channel estimation phase and combining during the downlink data transmission phase. By doing so, the \gls{bs} is able to acquire accurate instantaneous estimates of the effective channels, which are then used for efficient downlink precoding and result in a great enhancement of the network performance.

The contributions of this paper are summarized as follows.
\begin{itemize}
\item[$\bullet$] We present the novel concept of \gls{mimo} \name, which aims at designing a suitable shaping of the channel covariance matrices of the \glspl{ue} to enforce a full or partial separation of their signal subspaces that would be otherwise highly overlapping.
\item[$\bullet$] We point out that the exploitation of the non-Kronecker nature of massive \gls{mimo} channels is crucial to suitably alter the channel statistics perceived at the \gls{bs} by acting at the \gls{ue}-side.
\item[$\bullet$] We derive a tractable expression of the ergodic achievable sum rate under \gls{mimo} \name{} with the objective of characterizing the impact of the proposed framework on the system performance.
\item[$\bullet$] We optimize the \name{} strategies by minimizing the variance of the inter-\gls{ue} interference (as a metric to measure the spatial correlation) among the interfering \glspl{ue}. To this end, we propose a low-complexity block coordinate descent algorithm that is proved to converge to a limit point of the original nonconvex problem. For the two-UE case, this is shown to converge to a stationary point of the original problem.
\item[$\bullet$] We provide numerical results characterizing several scenarios where \gls{mimo} \name{} outperforms spatial multiplexing. In particular, a superior sum-rate performance is achieved when the spatial selectivity of the \gls{bs} is not sufficient to separate \glspl{ue} exhibiting high spatial correlation and the channel estimation is limited by strong pilot contamination.
\end{itemize}

\textit{Outline.} The rest of the paper is structured as follows. Section~\ref{sec:SM} introduces the system model for a general multi-\gls{ue} \gls{mimo} downlink system. Section~\ref{sec:CS} presents the concept of \gls{mimo} \name. Section~\ref{sec:CS_Opt} proposes efficient methods to optimize the \name{} vectors. Section~\ref{sec:NR} presents numerical results to evaluate the performance of the proposed framework. Finally, Section~\ref{sec:Con} concludes the paper.

\textit{Notation.} Lowercase and uppercase boldface letters denote vectors and matrices, respectively, whereas $(\cdot)^{\tran}$, $(\cdot)^{\herm}$, and $(\cdot)^{*}$ are the transpose, Hermitian transpose, and conjugate operators, respectively. $\| \cdot \|$ and $\| \cdot \|_{\rmF}$ represent the Euclidean norm for vectors and the Frobenius norm for matrices, respectively, whereas $\Exp[\cdot]$ and $\Var[\cdot]$ are the expectation and variance operators, respectively. $\I_{A}$ denotes the $A$-dimensional identity matrix and $\0$ represents the zero vector or matrix with proper dimension. $\tr(\cdot)$ and $\mathrm{vec}(\cdot)$ are the trace and vectorization operators, respectively. $(a_{mn})_{m,n=1}^M$ denotes the matrix of size $M\times M$ whose $(m,n)$th element is given by $a_{mn}$, whereas $[a_{1}, \ldots, a_{A}]$ represents horizontal concatenation. We use $\otimes$ to denote the Kronecker product and $\u_{\min}[\A]$ to denote the eigenvector corresponding to the minimum eigenvalue of matrix $\A$, with $\big\|\u_{\min}(\A)\big\| = 1$. $\mathbbm{1}_{A}$ denotes the indicator function, which is equal to $1$ if condition $A$ is satisfied and to $0$ otherwise, and $\ceil{\cdot}$ is the ceiling operator. Lastly, $\setC \setN(\0, \A)$ is the circularly symmetric complex Gaussian distribution with zero mean and covariance matrix $\A$.

\section{System Model} \label{sec:SM}

In this section, we introduce the channel model for a general multi-\gls{ue} \gls{mimo} downlink system. Then, we describe the uplink pilot-aided channel estimation assuming a \gls{tdd} setting and channel reciprocity between the uplink and the downlink. Finally, we discuss the system model for the downlink data transmission.

\subsection{Channel Model} \label{subsec:CM}

Consider a multi-\gls{ue} \gls{mimo} system where a \gls{bs} equipped with $M$ antennas serves $K$ \glspl{ue} with $N$ antennas each in the downlink. Let $\H_{k} \triangleq [\h_{k,1}, \ldots, \h_{k,M}] = [\g_{k,1}^{\tran}, \ldots, \g_{k,N}^{\tran}]^{\tran} \in \Compl^{N \times M}$ denote the downlink channel matrix of \gls{ue}~$k$, where $\h_{k,m} \in \Compl^{N \times 1}$ and $\g_{k,n} \in \Compl^{1 \times M}$ are the channel vectors between the $m$th \gls{bs} antenna and \gls{ue}~$k$ and between the \gls{bs} and the $n$th antenna of \gls{ue}~$k$, respectively. We assume a correlated Rayleigh fading channel model where the entries of $\H_{k}$ satisfy $\mathrm{vec} (\H_{k}) \sim \setC \setN (\0, \Sigmab_{k})$ \cite[Ch.~3]{Pau06}. Here, the channel covariance matrix $\Sigmab_{k} \in \Compl^{N M \times N M}$ has the following general structure:
\begin{align} \label{eq:Sigma_k}
\Sigmab_{k} \triangleq
\begin{bmatrix}
\Sigmab_{k,1 1}         & \Sigmab_{k,1 2} & \ldots & \Sigmab_{k,1 M} \\
\Sigmab_{k,1 2}^{\herm} & \Sigmab_{k,2 2} &        & \vdots \\
\vdots                  &                 & \ddots & \\
\Sigmab_{k,1 M}^{\herm} & \ldots          &        & \Sigmab_{k,M M}
\end{bmatrix}
\end{align}
where each block $\Sigmab_{k,m n} \triangleq \Exp [\h_{k,m} \h_{k,n}^{\herm}] \in \Compl^{N \times N}$ represents the cross-covariance matrix between the $m$th and $n$th columns of $\H_{k}$. Lastly, we define the covariance matrix seen at \gls{ue}~$k$ as $\R_{k} \triangleq \Exp [\H_{k} \H_{k}^{\herm}] \in \Compl^{N \times N}$ and the covariance matrix relative to \gls{ue}~$k$ seen at the \gls{bs} as $\T_{k} \triangleq \Exp [\H_{k}^{\herm} \H_{k}] \in \Compl^{M \times M}$, respectively. Observe that, in the case of downlink data transmission, $\R_{k}$ and $\T_{k}$ represent the receive and transmit covariance matrices, respectively.

\subsection{Uplink Pilot-Aided Channel Estimation} \label{subsec:UL_CE}

Assuming a \gls{tdd} setting and channel reciprocity between the uplink and the downlink (see, e.g., \cite[Ch.~1.3.5]{Bjo17}), the channel matrices $\{ \H_{k} \}_{k=1}^{K}$ are estimated at the \gls{bs} using antenna-specific uplink pilots, such that $N$ pilot vectors per \gls{ue} are required. Let $\setS_{p} \triangleq \{ k : \textrm{UE}~k~\textrm{has~pilot}~\P_{p} \}$ be the set of \glspl{ue} sharing the same pilot matrix $\P_{p} \in \Compl^{N \times \tau}$, with $p = 1, \ldots, P$. Here, $P<K$ denotes the number of orthogonal pilot matrices and $\tau$ represents the pilot length. The orthogonal pilot matrices satisfy $\{ \P_{p} \P_{p}^{\herm} = \frac{\tau}{N} \I_{N} \}_{p=1}^{P}$, i.e., $N$ orthogonal pilot vectors are assigned to each \gls{ue}, and $\{ \P_{p} \P_{q}^{\herm} = \0 \}_{p \neq q}.$ Note that these conditions imply $\tau \geq P N$. Furthermore, $\Y \in \Compl^{M \times \tau}$ denotes the receive signal at the \gls{bs} during the uplink pilot-aided channel estimation phase, which is given by
\begin{align}\label{eq:Y_p}
\Y \triangleq \sum_{p=1}^{P} \sum_{k \in \setS_{p}} \sqrt{\rho_{\rmUE}} \H_{k}^{\herm} \P_{p} + \Z
\end{align}
where $\rho_{\rmUE}$ is the transmit power at the \glspl{ue} and $\Z \in \Compl^{M \times \tau}$ is the noise term at the \gls{bs} with elements independently distributed  as $\setC \setN (0, \sigma_{\rmBS}^2)$. Let $\Phib_{k,n n} \triangleq \Exp [\g_{k,n}^{\tran} \g_{k,n}^{*}] \in \Compl^{M \times M}$ denote the covariance matrix of $\g_{k,n}$. Then, the \gls{mmse} estimate of $\g_{k,n}$, with $k \in \setS_{p}$, is given by (see, e.g., \cite[Ch.~3.2]{Bjo17})
\begin{align}\label{eq:g_hat_0}
\hat{\g}_{k,n}^{\herm} \triangleq \frac{N}{\tau\sqrt{\rho_{\rmUE}}} \Phib_{k,n n} \Q_{k,n n}^{-1}\Y \P_{p}^{\herm}\e_n \in \Compl^{M \times 1}
\end{align}
where $\e_n\in \Real^{N\times 1}$ is the $n$th column of $\I_{N}$, $\Q_{k,n n} \triangleq \big( \Phib_{k,n n} + \sum_{j \in \setS_{p}\setminus\{k\}} \Phib_{j,n n} + \frac{N}{\tau \varrho_{\rmUE}} \I_{M} \big) \in \Compl^{M \times M}$ is the normalized covariance matrix of the receive signal after correlation with $\P_{p}^{\herm}\e_n$, and $\varrho_{\rmUE} \triangleq \frac{\rho_{\rmUE}}{\sigma_{\rmBS}^2}$. Note that \eqref{eq:g_hat_0} can be seen as a superposition of channels estimated using the same pilot $\P_{p}^{\herm}\e_n$, which cannot be distinguished by the \gls{bs}: this phenomenon is referred to as pilot contamination \cite{Jos11}. Finally, the \gls{mmse} estimate of $\H_{k}$ is obtained as $\hat{\H}_{k} \triangleq [\hat{\g}_{k,1}^{\tran}, \ldots, \hat{\g}_{k,N}^{\tran}]^{\tran} \in \Compl^{N \times M}$.

\subsection{Downlink Data Transmission} \label{subsec:DL_data}

In the following, we focus on the downlink data transmission and assume that the \gls{bs} transmits $L_k \leq N$ independent symbols to \gls{ue}~$k$. We denote the transmit symbol vector for \gls{ue}~$k$ by $\s_{k} \in \Compl^{L_{k} \times 1}$, with $\Exp[\s_{k} \s_{k}^{\herm}] = \I_{L_{k}}$, and introduce the multi-\gls{ue} transmit symbol vector $\s \triangleq [\s_{1}^{\tran}, \ldots, \s_{K}^{\tran}]^{\tran} \in \Compl^{L \times 1},$ with $L \triangleq \sum_{k=1}^{K} L_{k}$. Before the transmission, the \gls{bs} precodes $\s$ using the multi-\gls{ue} precoding matrix $\W \triangleq [\W_{1}, \ldots, \W_{K}] \in \Compl^{M \times L},$ where $\W_{k} \triangleq [\w_{k,1}, \ldots, \w_{k,L_{k}}] \in \Compl^{M \times L_{k}}$ is the precoding matrix corresponding to $\s_{k}$ and $\W$ satisfies the power constrain $\| \W \|_{\rmF}^{2} = 1$. The receive signal at \gls{ue}~$k$ is thus given by
\begin{align} \label{eq:y_k}
\y_{k} & \triangleq \sqrt{\rho_{\rmBS}} \H_{k} \W_{k} \s_{k} + \sqrt{\rho_{\rmBS}} \sum_{j \neq k} \H_{k} \W_{j} \s_{j} + \z_{k} \in \Compl^{N \times 1}
\end{align}
where $\rho_{\rmBS}$ is the transmit power at the \gls{bs} and $\z_{k} \sim \setC \setN (0, \sigma_{\rmUE}^2\I_{N})$ is the noise term at the \glspl{ue}. Finally, \gls{ue}~$k$ decodes $\s_{k}$ as $\hat{\s}_{k} \triangleq \V_{k}^{\herm} \y_{k} \in \Compl^{L_{k} \times 1}$, where $\V_{k} \triangleq [\v_{k,1}, \ldots, \v_{k,L_{k}}] \in \Compl^{N \times L_{k}}$ is the corresponding combining matrix. Assuming perfect CSI at the UEs, the resulting sum rate is given by
\begin{align} \label{eq:R}
\! R \! \triangleq \! \sum_{k=1}^{K} \! \sum_{\ell=1}^{L_{k}} \log_{2} \! \bigg( \! 1 \! + \! \frac{|\v_{k, \ell}^{\herm} \H_{k} \w_{k, \ell}|^{2}}{\sum_{(j, s) \neq (k, \ell)} \! |\v_{k, \ell}^{\herm} \H_{k} \w_{j, s}|^{2} \! + \! \frac{1}{\varrho_{\rmBS}} \|\v_{k,\ell}\|^2} \! \bigg)
\end{align}
with $\varrho_{\rmBS} \triangleq \frac{\rho_{\rmBS}}{\sigma_{\rmUE}^2}$.

\section{Covariance Shaping at the UE-Side} \label{sec:CS}

Although \gls{mmse} precoding/combining can asymptotically remove any interference as $M$ grows large \cite{Bjo18}, this no longer holds in the presence of a large number of \glspl{ue} and finite \gls{bs} antennas. In particular, the number of \gls{bs} antennas needed to effectively separate interfering \glspl{ue} with \gls{mmse} precoding/combining increases with their spatial correlation, i.e., as the \glspl{ue} get closer to each other, and may become impractical in crowded scenarios. In this setting, the \gls{bs} can spatially separate signals corresponding to different \glspl{ue} and mitigate or eliminate  pilot contamination if their channel covariance matrices lie on orthogonal supports, i.e., if $\Sigmab_k \Sigmab_j = \0$ for a given pair of \glspl{ue}~$k$ and~$j$ (see, e.g., \cite{Bjo18}). However, the degree of statistical orthogonality among the \glspl{ue} is determined by their locations combined with the surrounding scattering environments and both these factors are generally beyond the control of network designers. Hence, signal subspace separation among the \glspl{ue} rarely occurs in practice.

In this context, we propose a novel method relying uniquely on statistical \gls{csi} and referred to as \textit{\gls{mimo} \name} (or simply \textit{\name}), which is applied at the \gls{ue}-side to enforce the aforementioned signal subspace separation in scenarios where the spatial selectivity of the \gls{bs} is not sufficient to separate the \glspl{ue}. While \name{} is conceived in such a way that it is agnostic to the operating frequency, we envision its application especially in sub-$6$~GHz frequency bands, where the availability of \gls{bs} antennas may be limited and further exacerbates the lack of sufficient spatial selectivity. According to \name, the \glspl{ue} preemptively apply a statistical beamforming vector, different for each \gls{ue}, that aims at spatially separating their transmissions, thus drastically reducing both pilot contamination and interference. Here, the original \gls{mimo} channel of each \gls{ue} is transformed into an effective \gls{miso} channel by combining the transmit/receive signal with the corresponding statistical beamforming vector. To this end, the knowledge of the channel covariance matrices of the \glspl{ue} is required. Although the time scale at which such information must be acquired and used to compute the \name{} vectors increases with the mobility of the \glspl{ue} (or, more generally, as the channel coherence time reduces), this still happens much less frequently than the instantaneous uplink pilot-aided channel estimation. Remarkably, the proposed method exploits the realistic non-Kronecker structure of massive \gls{mimo} channels that allows to suitably alter the channel statistics perceived at the \gls{bs} by acting at the \gls{ue}-side, thus turning a generally inconvenient model into a benefit.

Let $\v_{k} \in \Compl^{N \times 1}$ denote the statistical beamforming vector preemptively applied at \gls{ue}~$k$: in the rest of the paper, we refer to $\v_{k}$ as \textit{covariance shaping vector}.\footnote{Note that the \glspl{ue} can be configured for transmission with \name{} by the \gls{bs} in the same way as for codebook-based or non-codebook-based transmission in the current 5G \gls{nr} implementations (see, e.g., \cite[Ch.~11.3]{Dal18} for more details).} Hence, the original \gls{mimo} channel $\H_{k}$ between the \gls{bs} and each \gls{ue}~$k$ is transformed into the effective \gls{miso} channel $\bar{\g}_{k} \triangleq \v_{k}^{\herm} \H_{k} \in \Compl^{1 \times M}$. In this setting, it follows that $\bar{\g}_{k} \sim \setC \setN (\0, \bar{\Phib}_{k})$, where $\bar{\Phib}_{k} \in \Compl^{M \times M}$ is the effective channel covariance matrix defined as
\begin{align}
\bar{\Phib}_{k} & \triangleq \Exp [\bar{\g}_{k}^{\tran} \bar{\g}_{k}^{*}] \\
& = \big( (\I_{M} \otimes \v_{k}^{\herm}) \Sigmab_{k} (\I_{M} \otimes \v_{k}) \big)^{\tran}
\end{align}
with $\Sigmab_{k}$ introduced in \eqref{eq:Sigma_k} and where $\Exp [\|\bar{\g}_{k}\|^{2}] = \tr (\bar{\Phib}_{k})$. In the rest of this section, we describe how the two phases of uplink pilot-aided channel estimation and downlink data transmission are modified under \name{} and we provide a tractable expression of the resulting ergodic achievable sum rate. The optimization of the \name{} vectors is discussed in Section~\ref{sec:CS_Opt}. For ease of exposition, we focus on the case where \name{} is applied to all the \glspl{ue} in the systems. Nonetheless, the proposed method may be applied to one or more subsets of closely spaced \glspl{ue}, e.g., those who cannot be effectively separated with \gls{mmse} precoding/combining. In this context, \glspl{ue} adopting \name{} and others employing the multiple antennas for spatial multiplexing can be served simultaneously by the \gls{bs} in a transparent manner.

\begin{figure*}
\setcounter{equation}{15}
\begin{align}
\label{eq:sinr_ce}\gamma_k & = \frac{\tr (\bar{\Phib}_k \Q_k^{-1} \bar{\Phib}_k)^2}{\sum_{j=1}^{K} \tr (\bar{\Phib}_k \bar{\Phib}_j \Q_j^{-1} \bar{\Phib}_j) + \sum_{j \in \setS_{p} \setminus \{k\}} \tr (\bar{\Phib}_k \Q_j^{-1} \bar{\Phib}_j)^2 + \frac{1}{\varrho_{\rmBS}} \|\v_k\|^2 \sum_{j=1}^{K} \tr (\bar{\Phib}_j)}
\end{align}
\setcounter{equation}{7}
\hrulefill
\vspace{-2mm}
\end{figure*}

\subsection{Uplink Pilot-Aided Channel Estimation} \label{subsec:CS_UL_CE}

To estimate the effective channels $\{ \bar{\g}_{k} \}_{k=1}^{K}$ resulting from \name, the \gls{bs} assigns the same pilot vector $\p_{p} \in \Compl^{1 \times \tau}$ to all \glspl{ue} $k \in \setS_{p}$, with $\{ \|\p_{p}\|^{2} = \tau \}_{p=1}^{P}$ and $\{ \p_{p} \p_{q}^{\herm}=0 \}_{p \neq q}$. The receive signal at the \gls{bs} during the uplink pilot-aided channel estimation phase, which we denote by $\bar{\Y} \in \Compl^{M \times \tau}$, is given by (cf. \eqref{eq:Y_p})
\begin{align} \label{eq:Y_p_bar}
\bar{\Y} \triangleq \sum_{p=1}^{P} \sum_{k \in \setS_{p}} \sqrt{\rho_{\rmUE}} \bar{\g}_{k}^{\herm} \p_{p} + \Z.
\end{align}
Then, the \gls{mmse} estimate of $\bar{\g}_{k}$, with $k \in \setS_{p}$, is given by (cf. \eqref{eq:g_hat_0})
\begin{align}\label{eq:g_hat}
\hat{\bar{\g}}_{k}^{\herm} \triangleq \frac{1}{\tau\sqrt{\rho_{\rmUE}}} \bar{\Phib}_{k} \Q_k^{-1} \bar{\Y} \p_{p}^{\herm} \in \Compl^{M \times 1}
\end{align}
where $\Q_{k} \triangleq \bar{\Phib}_{k} + \sum_{j \in \setS_{p}\setminus\{k\}} \bar{\Phib}_{j} + \frac{1}{\tau\varrho_{\rmUE}} \I_M \in \Compl^{M\times M}$ is the normalized covariance matrix of the receive signal after correlation with $\p_{p}^{\herm}$. Note that the estimation of the effective channels only requires one pilot vector per \gls{ue}. Hence, the application of \name{} can potentially reduce the pilot length with respect to the estimation of the channel matrices $\{ \H_{k} \}_{k=1}^{K}$ described in Section~\ref{subsec:UL_CE}, which requires $N$ pilot vectors per \gls{ue}.

\subsection{Downlink Data Transmission} \label{subsec:CS_DL_data}

Focusing on the downlink data transmission, the \gls{bs} now transmits only one symbol $s_{k} \in \Compl$ to each \gls{ue}~$k$, i.e., $\{L_{k} = 1\}_{k=1}^{K}$. Hence, we have $\s = [s_{1}, \ldots, s_{K}]^{\tran} \in \Compl^{K \times 1}$ and the multi-\gls{ue} precoding matrix becomes $\W = [\w_{1}, \ldots, \w_{K}] \in \Compl^{M \times K}$, where $\w_{k} \in \Compl^{M \times 1}$ is the precoding vector corresponding to $s_{k}$. The receive signal at \gls{ue}~$k$ is thus given by (cf. \eqref{eq:y_k})
\begin{align} \label{eq:y_k_bar}
\bar{y}_{k} & \triangleq \sqrt{\rho_{\rmBS}} \bar{\g}_{k} \w_{k} s_{k} + \sqrt{\rho_{\rmBS}} \sum_{j \neq k} \bar{\g}_{k} \w_{j} s_{j} + \v_{k}^{\herm} \z_{k} \in \Compl
\end{align}
and, assuming perfect CSI at the UEs, the resulting sum rate is given by (cf. \eqref{eq:R})
\begin{align} \label{eq:R_bar}
\bar{R} \triangleq \sum_{k=1}^{K} \log_{2} \bigg( 1 + \frac{|\bar{\g}_{k} \w_{k}|^{2}}{\sum_{j \neq k} |\bar{\g}_{k} \w_{j}|^{2} + \frac{1}{\varrho_{\rmBS}} \|\v_{k}\|^{2}} \bigg).
\end{align}

\subsection{Ergodic Achievable Sum Rate} \label{subsec:ergodic_srate}

We now analyze the sum rate as a function of the \name{} vectors with the objective of characterizing the impact of the proposed framework on the system performance. In particular, we derive a tractable expression of the ergodic achievable sum rate by assuming that each \gls{ue}~$k$ does not know the effective scalar channel $\bar{\g}_k \w_k = \v_k^{\herm} \H_k \w_k$ instantaneously but only its expected value. Building on \cite[Thm.~4.6]{Bjo17}, we can express the receive signal in \eqref{eq:y_k_bar} as
\begin{align}
\bar{y}_k & = \sqrt{\rho_{\rmBS}} \Exp[\bar{\g}_k \w_k] s_k + z_k^{\prime}
\end{align}
where the effective noise term $z_k^{\prime} \triangleq \sqrt{\rho_{\rmBS}} \big( \bar{\g}_k \w_k - \Exp[\bar{\g}_k \w_k] \big) s_k + \sqrt{\rho_{\rmBS}} \sum_{j \neq k} \bar{\g}_k \w_j s_j + \v_k^{\herm} \z_k$ accounts for the lack of instantaneous \gls{csi} in addition to the interference and the noise at \gls{ue}~$k$. A lower bound on the \gls{sinr} achievable by UE~$k$ can be obtained by treating $z_{k}^{\prime}$ as Gaussian noise, which yields the effective \gls{sinr}
\begin{align} \label{eq:eff_sinr}
\gamma_k & \triangleq \frac{\big| \Exp[\bar{\g}_k \w_k] \big|^2}{\Var[\bar{\g}_k \w_k] + \sum_{j \neq k} \Exp \big[ |\bar{\g}_k \w_j|^2 \big] + \frac{1}{\varrho_{\rmBS}} \| \v_k \|^2}.
\end{align}

The expression in \eqref{eq:eff_sinr} can be further simplified by considering the case where the \gls{bs} adopts \gls{mrt} precoding. In this setting, the multi-\gls{ue} precoding matrix can be written as
\begin{align}
\W = \frac{\hat{\bar{\H}}^{\herm}}{\sqrt{\Exp[ \|\bar{\H}\|_{\mathrm{F}}^2]}}
\end{align}
where $\bar{\H} \! \triangleq \! [\bar{\g}_1^{\tran}, \ldots, \bar{\g}_K^{\tran}]^{\tran} \! \in \! \Compl^{K \times M}$ and $\hat{\bar{\H}} \! \triangleq \! [\hat{\bar{\g}}_1^{\tran}, \ldots, \hat{\bar{\g}}_K^{\tran}]^{\tran} \! \in \! \Compl^{K \times M}$ contain the effective channels and their \gls{mmse} estimates defined in \eqref{eq:g_hat}, respectively. Hence, when \gls{mrt} precoding is adopted at the \gls{bs}, the effective \gls{sinr} of \gls{ue}~$k$ in \eqref{eq:eff_sinr}, with $k \in \setS_{p}$, reduces to 
\begin{align}
\label{eq:sinr_ce_0} \gamma_k & = \frac{ \big| \Exp[\bar{\g}_k \hat{\bar{\g}}_k^{\herm}] \big|^2}{\Var[\bar{\g}_k \hat{\bar{\g}}_k^{\herm}] + \sum_{j \neq k} \Exp \big[ |\bar{\g}_k \hat{\bar{\g}}_j^{\herm}|^{2} \big] + \frac{1}{\varrho_{\rmBS}} \|\v_k\|^2 \Exp \big[ \|\bar{\H}\|_{\mathrm{F}}^2 \big]}.
\end{align}
This can be rewritten in closed form as in \eqref{eq:sinr_ce}, shown at the top of this page, by substituting the following terms: $\Exp[\bar{\g}_k \hat{\bar{\g}}_k^{\herm}] = \tr (\bar{\Phib}_k \Q_k^{-1} \bar{\Phib}_k)$, $\Var[\bar{\g}_k \hat{\bar{\g}}_k^{\herm}] = \tr (\bar{\Phib}_k^{2} \Q_k^{-1} \bar{\Phib}_k)$, $\Exp \big[ |\bar{\g}_k \hat{\bar{\g}}_j^{\herm}|^{2} \big] = \tr ( \bar{\Phib}_k \bar{\Phib}_j\Q_j^{-1}\bar{\Phib}_j)+ \mathbbm{1}_{j \in \setS_p}\tr (\bar{\Phib}_k\Q_j^{-1}\bar{\Phib}_j)^2$, and $\Exp \big[ \|\bar{\H}\|_{\mathrm{F}}^2 \big] = \sum_{k=1}^{K} \tr(\bar{\Phib}_k)$. We refer to Appendix~\ref{sec:app2} for the detailed derivations. Note that, in the case of perfect channel estimation, i.e., when $\varrho_{\rmUE} \to \infty$ and all the \glspl{ue} have orthogonal pilots, we have $\{ \bar{\Phib}_k\Q_k^{-1}\bar{\Phib}_k=\bar{\Phib}_k \}_{k=1}^{K}$ and $\big\{ \mathbbm{1}_{j \in \setS_p} = 0 \big\}_{j \neq k}$. In this context, the effective \gls{sinr} in \eqref{eq:sinr_ce} simplifies as \setcounter{equation}{16}
\begin{align}
\gamma_k & = \frac{\tr (\bar{\Phib}_k)^2}{\sum_{j=1}^{K} \tr (\bar{\Phib}_k \bar{\Phib}_j) + \frac{1}{\varrho_{\rmBS}} \|\v_k\|^2 \sum_{j=1}^{K} \tr (\bar{\Phib}_j)} \label{eq:sinr_perfcsit}.
\end{align}
Finally, Proposition~\ref{prop:achievable_sr} presents an ergodic achievable sum rate with MRT in the two cases of perfect and imperfect channel estimation. The proof follows from \cite[Thm.~4.6]{Bjo17} and is thus omitted.

\begin{proposition}\label{prop:achievable_sr}
Assume that the \gls{bs} adopts \gls{mrt} precoding. Then, an ergodic achievable sum rate is given by
\begin{align} \label{eq:R_bar_lb}
\bar{R}^{\lb} = \sum_{k=1}^K \log_{2}(1 + \gamma_k)
\end{align}
with $\gamma_k$ defined in \eqref{eq:sinr_ce} and in \eqref{eq:sinr_perfcsit} for imperfect and perfect channel estimation, respectively. Furthermore, the following limit holds as $M \to \infty$:
\begin{align}
\lim_{M \to \infty} \bar{R}^{\lb} = \Exp[\bar{R}]
\end{align}
with $\bar{R}$ defined in \eqref{eq:R_bar}.
\end{proposition}

\section{Covariance Shaping Optimization} \label{sec:CS_Opt}

In this section, we address the sum rate maximization through a proper design of the \name{} vectors at the \glspl{ue}. To this end, we consider the variance of the inter-\gls{ue} interference as a metric to measure the spatial correlation or, in other words, the degree of statistical orthogonality, between two interfering \glspl{ue}. In the following, we first consider the simple case of $K = 2$ and then extend the resulting analysis to the general case of $K \geq 2$.

Let us define the inter-\gls{ue} interference between \glspl{ue}~$k$ and~$j$ after applying \name{} as

\begin{align}
\Omega(\v_k, \v_j) \triangleq \frac{\bar{\g}_k(\v_k) \bar{\g}_j^{\herm}(\v_j)}{\sqrt{\Exp \big[ \| \bar{\g}_k(\v_k) \|^2 \big] \Exp \big[ \| \bar{\g}_j(\v_j) \|^2 \big]}}
\end{align}
where the notation $\bar{\g}_k(\v_k)$ makes explicit the dependence of the effective channel $\bar{\g}_k$ on the corresponding \name{} vector $\v_k$. The effective channels of \glspl{ue}~$k$ and~$j$ yield asymptotic \textit{favorable propagation} if they satisfy \cite[Ch.~2.5.2]{Bjo17}
\begin{align}
\lim_{M \to \infty} \Omega(\v_k, \v_j) = 0.
\end{align} 
For a practical number of \gls{bs} antennas $M$, a meaningful performance metric is the variance of $\Omega(\v_k, \v_j)$, expressed as
\begin{align}
\delta(\v_k,\v_j) & \triangleq \Var \big[ \Omega(\v_k, \v_j) \big] \\
& = \frac{\tr \big( \bar{\Phib}_k(\v_k) \bar{\Phib}_j(\v_j) \big)}{\tr \big( \bar{\Phib}_k(\v_k) \big) \tr \big( \bar{\Phib}_j(\v_j) \big)} \label{eq:delta_kj}
\end{align}
where the notation $\bar{\Phib}_k(\v_k)$ makes explicit the dependence of the effective channel covariance matrix $\bar{\Phib}_k$ on the corresponding \name{} vector $\v_k$. Note that the terms in the numerator and denominator of \eqref{eq:delta_kj} can be written as
\begin{align}
\tr \big( \bar{\Phib}_{k}(\v_{k}) \bar{\Phib}_{j}(\v_{j}) \big) & = \sum_{m,n=1}^{M} \v_{k}^{\herm} \Sigmab_{k,mn} \v_{k} \v_{j}^{\herm} \Sigmab_{j,nm} \v_{j} \\
& = \sum_{m,n=1}^{M} \v_{k}^{\herm} \Sigmab_{k,mn} \v_{k} \v_{j}^{\herm} \Sigmab_{j,mn}^{\herm} \v_{j}, \\
\tr \big( \bar{\Phib}_{k}(\v_{k}) \big) & = \v_{k}^{\herm} \bigg( \sum_{m=1}^{M} \Sigmab_{k,mm} \bigg) \v_{k}
\end{align}
where we recall that $\Sigmab_{k,m n}$ represents the cross-covariance matrix between the $m$th and $n$th columns of the original channel $\H_{k}$, i.e., before applying \name{}, as defined in \eqref{eq:Sigma_k}. Observe that $\delta(\v_k,\v_j)=0$ implies $\bar{\Phib}_k\bar{\Phib}_j=\0$, i.e.,  $\bar{\Phib}_k$ and $\bar{\Phib}_j$ lie on orthogonal supports \cite{Bjo18}. Indeed, considering the eigenvalue decomposition of the effective channel covariance matrices, which may be written as $\{\bar{\Phib}_i = \U_{i}\Lambdab_i\U_i^{\herm}\}_{i\in \{k,j\}}$, the condition of statistical orthogonality is satisfied if $\U_k = \U_j$ and $\tr(\Lambdab_k\Lambdab_j)=0,$ which implies the rank deficiency of both $\bar{\Phib}_k$ and $\bar{\Phib}_j$. Clearly, in the general case, this imposes $M^2$ conditions whereas only $2N$ variables can be adjusted: this means that the resulting system of equations can be solved when $N\geq \frac{M^2}{2}$, which is generally not verified in practice since $M \gg K N$ in massive \gls{mimo} scenarios. Hence, since full signal subspace separation can hardly be achieved by simple transceiver design at the \gls{ue}-side, it is of interest to minimize the spatial correlation between each pair of interfering \glspl{ue}.

\subsection{Two-UE Case} \label{subsec:2UE}

In the two-\gls{ue} case, i.e., for $K=2$, the \name{} vectors of \glspl{ue}~$k$ and~$j$ are computed by solving the optimization problem
\begin{align} \label{eq:P1} \tag{P1}
\begin{array}{cl}
\displaystyle \minimize_{\v_{k}, \v_j} & \displaystyle \delta(\v_k,\v_j) \\
\subjectto & \displaystyle \| \v_{k} \|^{2} = \| \v_{j} \|^{2} = 1
\end{array}
\end{align}
with $\delta(\v_k,\v_j)$ defined in \eqref{eq:delta_kj}. Although problem~\eqref{eq:P1} is not convex in either $\v_{k}$ or $\v_j$, a suboptimal solution can be efficiently obtained via alternating optimization, as suggested in \cite{Mur18}. Let us introduce the definition
\begin{align} \label{eq:etaj}
\eta_{j,mn}(\v_{j}) \triangleq \frac{\v_{j}^{\herm} \Sigmab_{j,mn}^{\herm} \v_{j}}{\v_{j}^{\herm} \big( \sum_{m=1}^{M} \Sigmab_{j,mm} \big) \v_{j}}.
\end{align}
The optimal \name{} vector of \gls{ue}~$k$ for a given $\v_j$, denoted by $\v_k^{\star}$, is obtained as
\begin{align} \label{eq:P1_1}
\v_{k}^{\star} = \argmin_{\v_{k} \, : \, \| \v_{k} \|^{2} = 1} \frac{\sum_{m,n=1}^{M} \v_{k}^{\herm} \big( \eta_{j,mn}(\v_{j}) \Sigmab_{k,mn} \big) \v_{k}}{\v_{k}^{\herm} \big( \sum_{m=1}^{M} \Sigmab_{k,mm} \big) \v_{k}}.
\end{align}
Since \eqref{eq:P1_1} is in the form of generalized Rayleigh quotient, it admits the solution
\begin{align} \label{eq:vk_star}
\v_{k}^{\star} = \u_{\min}\bigg[\bigg( \sum_{m=1}^{M} \Sigmab_{k,mm}\bigg)^{-1}\bigg(\sum_{m,n=1}^{M}\eta_{j,mn}(\v_{j}) \Sigmab_{k,mn}\bigg)\bigg]
\end{align}
and the optimal \name{} vector of \gls{ue}~$j$ for a given $\v_k$ is obtained in a similar way. Hence, problem~\eqref{eq:P1} is solved by alternating the optimization between $\v_k$ and $\v_j$ until a predetermined convergence criterion is satisfied, e.g., until the difference between the values of the objective in consecutive iterations is sufficiently small. This scheme is formalized in Algorithm~\ref{alg:A1}, whose convergence properties are characterized in Proposition~\ref{prop:prop_2UE}.

\begin{figure}[t!]
\vspace{-3mm}
\begin{algorithm}[H]
\begin{algorithmic}
\STATE \hspace{-4mm} \textbf{Data:} $\v_k^{(0)}$, $\v_j^{(0)}$, $\Sigmab_{k}$, $\Sigmab_{j}$, and the accuracy $\epsilon$. Fix $n=1$.
\STATE \hspace{-4mm} \texttt{While} $\big| \delta(\v_k^{(n)},\v_j^{(n)}) - \delta(\v_k^{(n-1)},\v_j^{(n-1)}) \big|/\delta(\v_k^{(n)},\v_j^{(n)}) > \epsilon$
\begin{itemize}[leftmargin=12mm]
\item[\texttt{S.1:}] Given $\v_j^{(n-1)}$, compute $\{\eta_{j,mn}(\v_{j}^{(n-1)})\}_{m,n=1}^{M}$ as in \eqref{eq:etaj}.
\item[\texttt{S.2:}] Compute $\v_k^{(n)}$ as in \eqref{eq:vk_star}.
\item[\texttt{S.3:}] Given $\v_k^{(n)}$, compute $\{\eta_{k,mn}(\v_{k}^{(n)})\}_{m,n=1}^{M}$ as in \eqref{eq:etaj}.
\item[\texttt{S.4:}] Compute $\v_j^{(n)}$ as in \eqref{eq:vk_star}.
\end{itemize}
\STATE \hspace{-4mm} \texttt{End}
\begin{itemize}[leftmargin=7.5mm]
\item[\texttt{S.5:}] Fix $\v_k = \v_k^{(n)}$ and $\v_j = \v_j^{(n)}$.
\end{itemize}
\vspace{-1mm}
\end{algorithmic}
\caption{(\Name: Alternating optimization algorithm)} \label{alg:A1}
\end{algorithm}
\vspace{-5mm}
\end{figure}

\begin{proposition} \label{prop:prop_2UE}
The alternating optimization algorithm described in Algorithm~\ref{alg:A1} converges to a stationary point of problem~\eqref{eq:P1}.
\end{proposition}

\begin{IEEEproof}
Observe that the objective function in \eqref{eq:P1_1} does not depend on the scaling of $\v_k$ and, if we relax the nonconvex constraint $\|\v_k\|^2=1$ as $\|\v_k\|^2\leq 1$, \eqref{eq:vk_star} remains a solution of \eqref{eq:P1_1}. Then, according to \cite[Corollary 2]{Gri00}, every limit point of the sequence generated by the alternating optimization algorithm applied to problem~\eqref{eq:P1} with the relaxed constraints is a stationary point of the original problem.
\end{IEEEproof} \vspace{1mm}

Algorithm~\ref{alg:A1} can be implemented in a centralized manner at the \gls{bs} and the \name{} vectors are fed back to the corresponding \glspl{ue}. Alternatively, it can be implemented in a distributed fashion at the \glspl{ue}. In this case, each \gls{ue} can compute its \name{} vector without any information exchange with the other \gls{ue} provided that the channel statistics of the latter are known, the order of update is fixed, and the same initial points are used.\footnote{Note that $\v_k^{(0)}$ and $\v_j^{(0)}$ can be any predefined pair of normalized vectors provided by the \gls{bs}.} In fact, under these conditions, the entire alternating optimization procedure can be carried out locally and independently at each \gls{ue}. Note that the same considerations hold for Algorithm~\ref{alg:A2} presented in the next section for the multi-\gls{ue} case, where we also discuss the computational complexity.

\begin{figure}[t]
\centering
\begin{subfigure}{0.45\textwidth}
\centering
\includegraphics[scale=0.8]{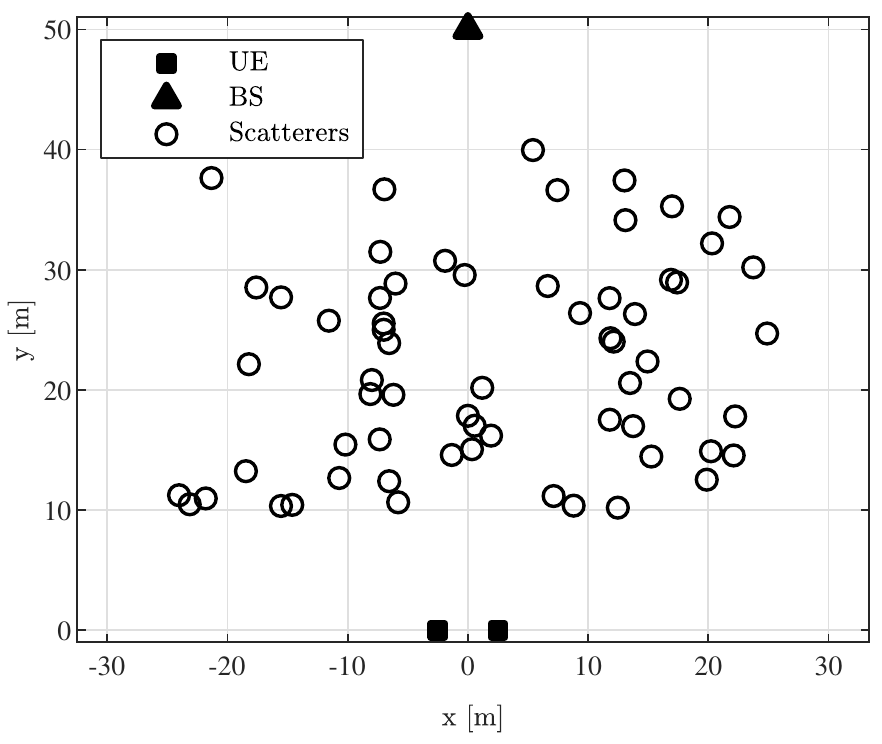}
\caption{NLoS scenario with $K=2$.}
\label{fig:map_2UE_nlos}
\end{subfigure} \hfill
\begin{subfigure}{0.45\textwidth}
\centering
\vspace{3.8mm}
\hspace{4.6mm} \includegraphics[scale=0.85]{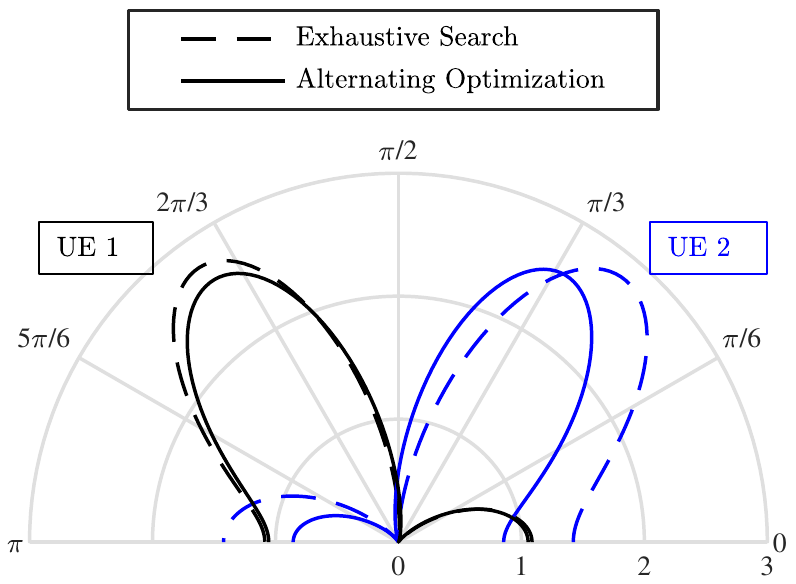}
\vspace{0.6mm}
\caption{Corresponding \name{} vectors.}
\label{fig:v_vectors_2UE_nlos}
\end{subfigure}
\caption{Two-\gls{ue} case: (a) 2D map of the considered NLoS scenario with inter-\gls{ue} distance $d=4$~m; (b) corresponding \name{} vectors with $N=2$ obtained with exhaustive search and with Algorithm~\ref{alg:A1}.} \label{fig:2} \vspace{-3mm}
\end{figure}

\begin{figure}[t]
\centering
\begin{subfigure}{0.45\textwidth}
\centering
\includegraphics[scale=0.8]{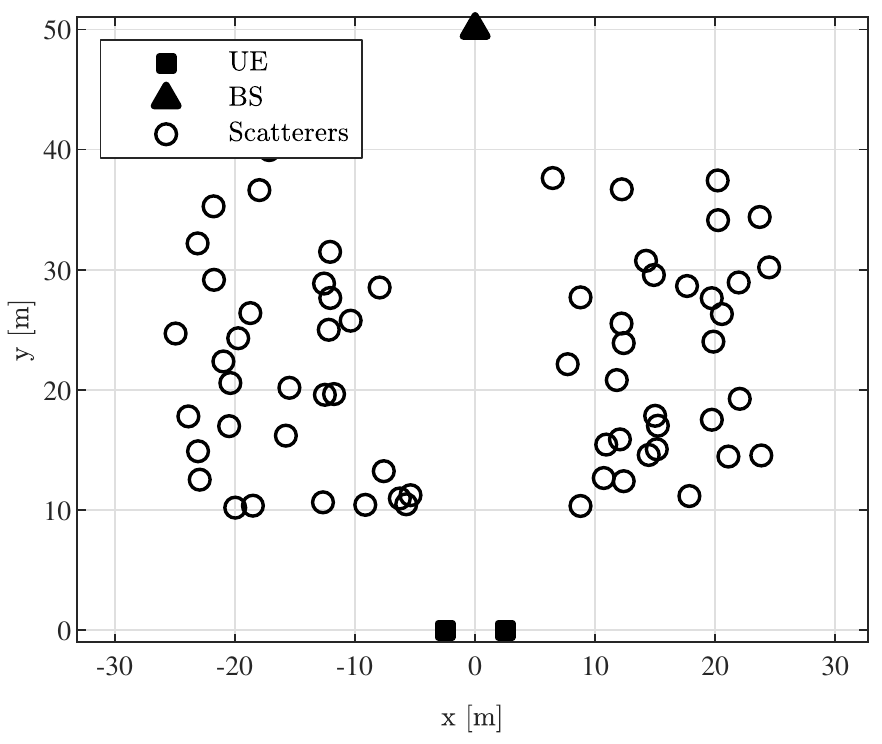}
\caption{LoS scenario with $K=2$.}
\label{fig:map_2UE_los}
\end{subfigure} \hfill
\begin{subfigure}{0.45\textwidth}
\centering
\vspace{4.7mm}
\hspace{4.6mm} \includegraphics[scale=0.85]{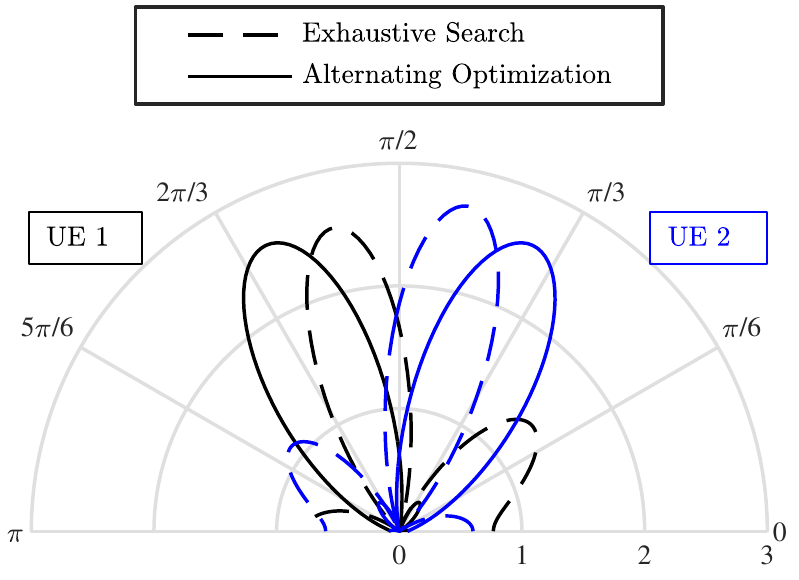}
\caption{Corresponding \name{} vectors.}
\label{fig:v_vectors_2UE_los}
\end{subfigure}
\caption{Two-\gls{ue} case: (a) 2D map of the considered LoS scenario with inter-\gls{ue} distance $d=4$~m; (b) corresponding \name{} vectors with $N=2$ obtained with exhaustive search and with Algorithm~\ref{alg:A1}.} \label{fig:3} \vspace{-3mm}
\end{figure}

The result of the \name{} optimization heavily depends on the physical scattering environment. Consider the \gls{nlos} scenario in Fig.~\ref{fig:2}(a), where there is no \gls{los} path between the \gls{bs} and the \glspl{ue}, the \glspl{ue} are equipped with \glspl{ula}, and the inter-\gls{ue} distance is $d=4$~m. In this case, as shown in Fig.~\ref{fig:2}(b), the \name{} vectors tend to focus their power along reflected paths that interfere with each other as little as possible while also carrying sufficient channel power, which results in a nearly interference-free transmission/reception. On the other hand, when a \gls{los} path exists, as in the \gls{los} scenario in Fig.~\ref{fig:3}(a), this generally carries more channel power than any other path. In this case, as shown in Fig.~\ref{fig:3}(b), the propagation directions selected by the \name{} vectors tend to partially capture the \gls{los} path while also focusing some of their power along separated reflected paths in order to achieve some degree of statistical orthogonality between the \glspl{ue}. Remarkably, in both Fig.~\ref{fig:2}(b) and Fig.~\ref{fig:3}(b), the result of Algorithm~\ref{alg:A1} is very close to the optimal solution of problem~\eqref{eq:P1} obtained with exhaustive search within a set of $10^3$ uniformly distributed vectors over the $N$-dimensional unit sphere for each \gls{ue}, where the former is characterized by negligible complexity with respect to the latter.

\vspace{1mm}

\noindent \textbf{Kronecker Channel Model.} Let us consider the particular case where each channel $\H_{k}$ follows the Kronecker channel model \cite{Wei06}. In this setting, we have $\H_{k} =\R_{k}^{\frac{1}{2}} \H_{k}^{(\rmw)} \T_{k}^{\frac{1}{2}}$, with $\R_{k}$ and $\T_{k}$ defined in Section~\ref{sec:SM} and $\mathrm{vec} (\H_{k}^{(\rmw)}) \sim \setC \setN (\0, \I_{N M})$. Accordingly, the channel covariance matrix in \eqref{eq:Sigma_k} can be expressed as $\Sigmab_{k} = \T_{k}^{\tran} \otimes \R_{k}$ with block elements given by $\Sigmab_{k,m n} = T_{k,m n}^{*} \R_{k}$, where $T_{k,m n}$ denotes the $(m,n)$th element of $\T_{k}$. Hence, from \eqref{eq:delta_kj}, we have

\begin{align}
& \delta (\v_{k}, \v_{j}) \nonumber \\
& = \frac{\tr \big( (\v_{k}^{\herm} \Sigmab_{k,m n} \v_{k})_{m,n=1}^{M} (\v_{j}^{\herm} \Sigmab_{j,m n}^{\herm} \v_{j})_{m,n=1}^{M} \big)}{\tr\big((\v_{k}^{\herm} \Sigmab_{k,m n} \v_{k})_{m,n=1}^{M} \big)\tr\big((\v_{j}^{\herm} \Sigmab_{j,m n}^{\herm} \v_{j})_{m,n=1}^{M} \big)} \\
& = \frac{\tr \big( (T_{k,m n} \v_{k}^{\herm} \R_{k} \v_{k})_{m,n=1}^{M}(T_{j,m n} \v_{j}^{\herm} \R_{j} \v_{j})_{m,n=1}^{M} \big)}{\tr \big( (T_{k,m n} \v_{k}^{\herm} \R_{k} \v_{k} )_{m,n=1}^{M}\big)\tr\big((T_{j,m n}\v_{j}^{\herm} \R_{j} \v_{j})_{m,n=1}^{M}\big)} \\
& = \frac{\tr (\T_{k} \T_{j})}{\tr(\T_k)\tr(\T_j)}.
\end{align}
It is straightforward to observe that, in this case, $\delta (\v_{k}, \v_{j})$ is independent of $\v_{k}$ and $\v_{j}$. Hence, under the Kronecker channel model, it is not possible to alter the channel statistics perceived at one end of the communication link by designing the transceiver at the other end: as a consequence, no meaningful effective channel separation can be performed. This is in accordance with the properties of the Kronecker channel model, whereby the transmit and receive covariance matrices are independent. In this context, the signal subspace separation is exclusively determined by the physical scattering environment and can only be achieved when $\tr (\T_{k} \T_{j}) = 0$, which is rarely satisfied in practice \cite{Rag10,Wei06}.

\subsection{Multi-UE Case} \label{subsec:MultiUE}

In the general case, i.e., for $K \geq 2$, the \name{} vectors for each \gls{ue} are computed by solving the optimization problem
\begin{align} \label{eq:P2} \tag{P2}
\begin{array}{cl}
\displaystyle \minimize_{\{ \v_k \}_{k=1}^K} & \displaystyle \sum_{\substack{k\neq j}} \delta(\v_k, \v_j) \\
\subjectto & \displaystyle \big\{ \| \v_{k} \|^{2} = 1 \big\}_{k=1}^{K}
\end{array}
\end{align}
with $\delta(\v_k,\v_j)$ defined in \eqref{eq:delta_kj}. Although problem~\eqref{eq:P2} is not convex in any of the optimization variables $\{ \v_k \}_{k=1}^K$, a suboptimal solution can be efficiently obtained via block coordinate descent, which can be interpreted as an extension of the alternating optimization approach presented in the previous section for the two-\gls{ue} case. The optimal \name{} vector of \gls{ue}~$k$ for given $\{\v_j\}_{j\neq k}$ is obtained as
\begin{align} \label{eq:P2_2}
\v_{k}^{\star} = \argmin_{\v_{k} \, : \, \| \v_{k} \|^{2} = 1} & \frac{\sum_{m,n=1}^{M} \v_{k}^{\herm} \big( \sum_{j\neq k} \eta_{j,mn}(\v_{j}) \Sigmab_{k,mn} \big) \v_{k}}{\v_{k}^{\herm} \big( \sum_{m=1}^{M} \Sigmab_{k,mm} \big) \v_{k}}
\end{align}
with $\eta_{j,mn}(\v_{j})$ defined in \eqref{eq:etaj}. Like \eqref{eq:P1_1}, \eqref{eq:P2_2} is in the form of generalized Rayleigh quotient and thus admits the solution
\begin{align}
\v_k^{\star} = \u_{\min} \! \bigg[ \bigg( \sum_{m=1}^{M} \Sigmab_{k,mm} \! \bigg)^{-1} \! \bigg( \sum_{m,n=1}^{M} \sum_{j\neq k} \eta_{j,mn} (\v_{j}) \Sigmab_{k,mn} \! \bigg) \bigg] \label{eq:vk_star2}
\end{align}
and the optimal \name{} vectors of the other \glspl{ue}~$j$ for given $\{\v_k\}_{k\neq j}$ are obtained in a similar way. Hence, problem~\eqref{eq:P2} is solved by optimizing the strategy of each \gls{ue} given the strategies of the other \glspl{ue} until a predetermined convergence criterion is satisfied. Furthermore, at each iteration~$i$, the update $\v_k^{(i)} = \alpha \v_k^{\star} + (1-\alpha) \v_k^{(i-1)}$ can be used to limit the variation of the \name{} vectors between consecutive iterations, where the step size $\alpha \in (0,1]$ is chosen to strike the proper balance between convergence speed and accuracy (see, e.g., \cite{Atz21,Scu14}). The proposed scheme is formalized in Algorithm~\ref{alg:A2}, whose convergence properties are characterized in Proposition~\ref{prop:prop_gen}.

\begin{proposition} \label{prop:prop_gen}
The block coordinate descent algorithm described in Algorithm~\ref{alg:A2} converges to a limit point of problem~\eqref{eq:P2}.
\end{proposition}

\begin{figure}[t!]
\vspace{-3mm}
\begin{algorithm}[H]
\begin{algorithmic}
\STATE \hspace{-4mm} \textbf{Data:} $\{\v_k^{(0)}\}_{k=1}^K$, $\{ \Sigmab_{k} \}_{k=1}^K$, $\alpha \in (0,1]$, and the accuracy $\epsilon$. Fix $n=1$.
\STATE \hspace{-4mm} \texttt{While} $$\sum_{\substack{k\neq j}} \frac{\big| \delta(\v_k^{(n)}, \v_j^{(n)}) - \delta(\v_k^{(n-1)}, \v_j^{(n-1)}) \big|}{\delta(\v_k^{(n)}, \v_j^{(n)})} > \epsilon$$
\STATE \texttt{For} $k=1,\ldots,K$
\begin{itemize}[leftmargin=16mm]
\item[\texttt{S.1}:] Given $\{\v_j^{(n-1)}\}_{j\neq k}$, compute\\$\{\eta_{j,mn}(\v_{j}^{(n)})\}_{m,n =1}^M$ as in \eqref{eq:etaj}, $\forall j \neq k$.
\item[\texttt{S.2}:] Compute $\v_k^{\star}$ as in \eqref{eq:vk_star2}.
\item[\texttt{S.3}:] Update $\v_k^{(n)} = \alpha \v_k^{\star} + (1-\alpha)\v_k^{(n-1)}$.
\end{itemize}
\STATE \texttt{End}
\STATE \hspace{-4mm} \texttt{End}
\begin{itemize}[leftmargin=7.5mm]
\item[\texttt{S.4}:] Fix $\{ \v_k = \v_k^{(n)} \}_{k=1}^K$.
\end{itemize}
\vspace{-1mm}
\end{algorithmic}
\caption{(\Name: Block coordinate descent algorithm)} \label{alg:A2}
\end{algorithm}
\vspace{-5mm}
\end{figure}

\begin{IEEEproof}
At each iteration~$i$ of Algorithm~\ref{alg:A1}, the \name{} vector of \gls{ue}~$k$ results from solving \eqref{eq:P1_1}, which admits the optimal solution in \eqref{eq:vk_star}. Hence, the sequence $\big\{ \delta(\v_k^{(i)},\v_j^{(i)}) \big\}_{i}$ is non-increasing since
\begin{align}
\delta(\v_k^{(i)},\v_j^{(i-1)}) \leq \delta(\v_k^{(i-1)},\v_j^{(i-1)}).
\end{align}
Moreover, since $\delta(\v_k,\v_j)\geq 0$, the sequence $\big\{ \delta(\v_k^{(i)},\v_j^{(i)}) \big\}_{i}$ converges to a finite non-negative value.
\end{IEEEproof}

\vspace{1mm}

The computational complexity of Algorithm~\ref{alg:A2} is essentially dictated by step~\texttt{S.2}, which corresponds to the computation of the optimal \name{} vector of \gls{ue}~$k$ for given $\{\v_j\}_{j\neq k}$. This is expressed in \eqref{eq:vk_star2} and has a complexity $\setO(N^{2.37})$ due to the eigenvalue decomposition and inverse matrix operations \cite{Yin16}. Hence, the overall computational complexity of Algorithm~\ref{alg:A2} is $\setO(I K N^{2.37})$, where $I$ is the number of iterations required for convergence. Note that both $N$ and $I$ are fairly modest since the \glspl{ue} are equipped with a small-to-moderate number of antennas and the algorithm converges in very few iterations (i.e., less than $10$ in the simulation scenarios considered in Section~\ref{sec:NR}). We can thus conclude that Algorithms~\ref{alg:A1} and~\ref{alg:A2} are remarkably efficient in terms of computation complexity.

\section{Numerical Results and Discussion} \label{sec:NR}

In this section, we present numerical results to evaluate the gains achieved by the proposed \name{} framework. To this end, we compare the following alternative transmission/reception schemes, where the first is based on \name{} and the second is considered as a reference.\footnote{The relevant spatial multiplexing methods in \cite{Spe04,Chr08} were also tested as reference schemes. Nonetheless, they are not included in this section since they are always outperformed by the reference scheme in~(B) in the considered simulation scenarios.}
\begin{itemize}
\item[(A)] \textbf{\Name.} The multiple antennas at the \gls{ue} are employed to implement \name{} and one data stream per \gls{ue} is transmitted using \gls{mmse} precoding. Specifically, each \gls{ue}~$k$ applies its \name{} vector $\v_k$, obtained with Algorithm~\ref{alg:A1} or Algorithm~\ref{alg:A2} (depending on the value of $K$), during both the uplink pilot-aided channel estimation phase and the downlink data transmission phase, as discussed in Section~\ref{sec:CS_Opt}. The \gls{bs} obtains the \gls{mmse} estimates of the resulting effective channels $\{\bar{\g}_k\}_{k=1}^{K}$ as in \eqref{eq:g_hat} based on the transmission of $P<K$ orthogonal pilot vectors by the \glspl{ue}. These estimates are used to compute the $(M \times K)$-dimensional \gls{mmse} precoding matrix at the \gls{bs} as in \cite{Bjo18}.
\item[(B)] \textbf{Spatial multiplexing.} The multiple antennas at the \gls{ue} are employed for spatial multiplexing and $N$ data streams per \gls{ue} are transmitted using \gls{mmse} precoding. Specifically, the \gls{bs} obtains the \gls{mmse} estimates of the channel matrices $\{\H_k\}_{k=1}^{K}$ as in \eqref{eq:g_hat_0} based on the transmission of $P<K$ orthogonal pilot matrices by the \glspl{ue}. These estimates are used to compute the $(M \times K N)$-dimensional \gls{mmse} precoding matrix at the \gls{bs} as in \cite{Bjo18}.
\end{itemize}
Some comments are in order.
\begin{itemize}
\item[$\bullet$] The reference scheme in~(B) requires antenna-specific pilots for the estimation of the channel matrices $\{ \H_{k} \}_{k=1}^{K}$. In particular, $N$ orthogonal pilot vectors are assigned to each \gls{ue} to avoid any pilot contamination among the antennas of the same \gls{ue}, which implies $\tau \geq P N$. On the other hand, the estimation of the effective channels $\{ \bar{\g}_{k} \}_{k=1}^{K}$ resulting from \name{} requires $\tau \geq P$ and, therefore, the pilot length can potentially be reduced.
\item[$\bullet$] The reference scheme in~(B) allows the transmission of up to $N$ data streams per \gls{ue}, while only one data stream per \gls{ue} is transmitted when \name{} is used. In the following, we demonstrate how the proposed \name{} method can effectively outperform spatial multiplexing in scenarios where the \glspl{ue} exhibit highly overlapping channel covariance matrices and the spatial selectivity of the \gls{bs} is not sufficient to separate such \glspl{ue}.
\end{itemize}

We evaluate different scenarios in which the \gls{bs} serves $K$ closely spaced \glspl{ue}. We assume that the \glspl{ue} are equipped with \glspl{ula}, whereas both \glspl{ula} and \glspl{upa} are considered at the \gls{bs}. In this context, we define the \gls{ula} response at each \gls{ue} for the angle of impingement $\theta$ as
\begin{align}
\a(\theta) & \triangleq [1, e^{-2\pi\delta\cos(\theta)}, \ldots, e^{-2\pi\delta(N-1)\cos(\theta)}]^{\tran} \in \Compl^{N \times 1}
\end{align}
where $\delta=0.5$ represents the ratio between the antenna spacing and the signal wavelength. Likewise, we define the \gls{upa} response at the \gls{bs} for the angles of impingement $\phi$ along the azimuth direction and $\psi$ along the elevation direction as
\begin{align}
& \hspace{-1.5mm} \b(\phi,\psi) \nonumber \\
& \hspace{-1.5mm} \triangleq [1, e^{-2\pi\delta\cos(\phi)}, \ldots, e^{-2\pi\delta(M_x-1)\cos(\phi)}]^{\tran} \nonumber \\
& \hspace{-1.5mm} \phantom{=} \ \otimes [1, e^{-2\pi\delta\sin(\psi)}, \ldots, e^{-2\pi\delta(M_y-1)\sin(\psi)}]^{\tran} \in \Compl^{M \times 1} \label{eq:pla_b}
\end{align}
where $M_x$ and $M_y$ represent the number of \gls{bs} antennas along the azimuth and elevation directions, respectively, with $M=M_xM_y$. The \gls{ula} response at the \gls{bs} for the angle of impingement $\phi$ can be recovered from \eqref{eq:pla_b} by setting $M_x=M$, $M_y=1$, and $\psi=0$. Then, the instantaneous channel between the \gls{bs} and each \gls{ue}~$k$ follows the discrete physical channel model in \cite{Say02} and is given by
\begin{align}
\H_k & = \sqrt{\frac{\kappa}{1+\kappa}} d_{k}^{-\frac{\beta}{2}} \a(\theta_{k})\b(\phi_{k},\psi_{k})^{\herm}\nonumber \\
& \phantom{=} \ + \sqrt{\frac{1}{1+\kappa}}\sum_{u=1}^{U} d_{k,u}^{-\frac{\beta}{2}} \alpha_{k,u} \a(\theta_{k,u})\b(\phi_{k,u},\psi_{k,u})^{\herm}
\end{align}
where $\kappa$ is the Ricean factor, $U$ is the number of reflected paths, $d_k$ and $d_{k,u}$ are the distances of the \gls{los} path and of the $u$th reflected path, respectively, $\theta_k$ and $\theta_{k,u}$ are the angles of impingement at the \gls{ue} of the \gls{los} path and of the $u$th reflected path, respectively, $\phi_k$ (resp. $\psi_k$) and $\phi_{k,u}$ (resp. $\psi_{k,u}$) are the angles of impingement at the \gls{bs} of the \gls{los} path and of the $u$th reflected path along the azimuth (resp. elevation) direction, respectively, $\alpha_{k,u}$ is the random phase delay of the $u$th reflected path, and $\beta$ is the pathloss exponent. Unless otherwise stated, we use the simulation parameters listed in Table~\ref{tab:param}. The following results are obtained by means of Monte Carlo simulations with $5 \times 10^{3}$ independent channel realizations. Lastly, we point out that, in the considered simulation scenarios, Algorithms~\ref{alg:A1} and~\ref{alg:A2} are always observed to converge in very few iterations, i.e., less than $10$.

\begin{table}[t!]
\centering
\small
\renewcommand*{\arraystretch}{1.25}
\begin{tabular}{llccc}
\hline
\textbf{Parameter}                  & & \textbf{Symbol}     & & \textbf{Value} \\
\hline
Number of \gls{bs} antennas         & & $M$                 & & $128$ \\
Number of \glspl{ue}                & & $K$                 & & $\{2,4,8\}$ \\
Number of \gls{ue} antennas         & & $N$                 & & $2$ \\
Noise variance at the \gls{bs}      & & $\sigma_{\rmBS}^2$  & & $-80$~dBm \\
Noise variance at the \glspl{ue}    & & $\sigma_{\rmUE}^2$  & & $-80$~dBm \\
Transmit power at the \gls{bs}      & & $\rho_{\rmBS}$      & & $30$~dBm \\
Transmit power at the \glspl{ue}    & & $\rho_{\rmUE}$      & & $25$~dBm \\
Number of orthogonal pilots         & & $P$                 & & $\big\lceil \frac{K}{2} \big\rceil$ \\
Pilot length                        & & $\tau$              & & $\{1,2,4\}$ \\
Ricean factor                       & & $\kappa$            & & \makecell{$2.5$ (\gls{los}) \\ $0$ (\gls{nlos})} \\
Pathloss exponent                   & & $\beta$             & & $2$ \\
Inter-\gls{ue} distance             & & $d$                 & & $\{2,4\}$~m \\
\hline
\end{tabular}
\vspace{1mm}
\caption{Simulation parameters (unless otherwise stated).} \label{tab:param}
\vspace{-3mm}
\end{table}

\subsection{Two-UE case} \label{subsec:NR_2UE}

Following the discussion in Section~\ref{subsec:2UE}, we first study the simple case of $K=2$ \glspl{ue} in order to highlight the key features of the proposed \name{} framework. In this setting, we assume $P=1$, i.e., the \glspl{ue} are assigned the same pilot, and we remark that there is no pilot contamination among the antennas of the same \gls{ue} for the reference scheme in~(B). Assuming that the \gls{bs} is equipped with a \gls{ula}, we consider the \gls{los} and \gls{nlos} scenarios illustrated in Fig.~\ref{fig:2}(a) and in Fig.~\ref{fig:3}(a), respectively, where the corresponding \name{} vectors are shown in Fig.~\ref{fig:2}(b) and in Fig.~\ref{fig:3}(b), respectively. Observe that, in both cases, the inter-\gls{ue} distance is $d=4$~m: this is quite small compared with the distance of about $50$~m between the \glspl{ue} and the \gls{bs}, and results in highly overlapping channel covariance matrices.

\begin{figure}[t]
\centering
\includegraphics[scale=0.625]{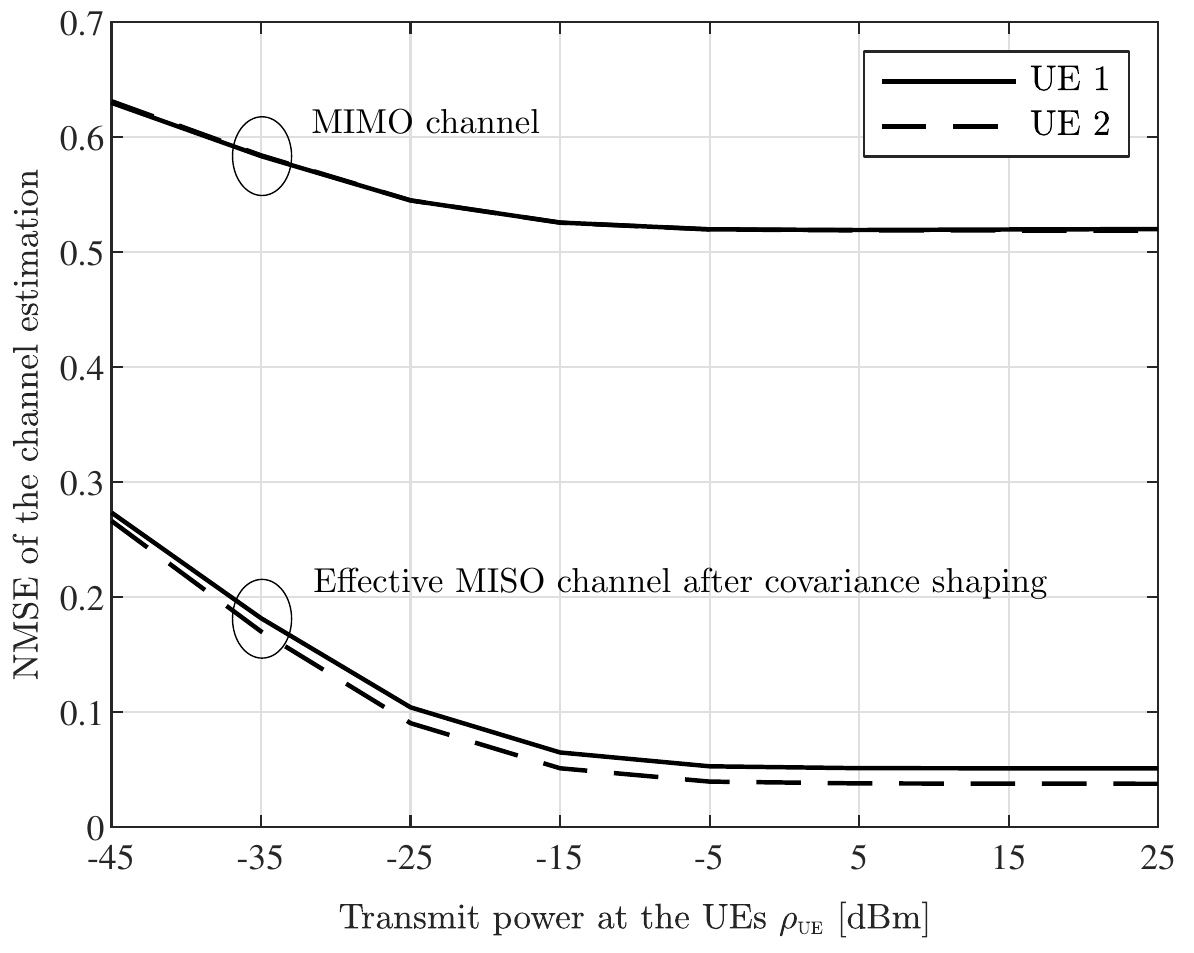}
\caption{Two-\gls{ue} case: \gls{nmse} of the channel estimation versus the transmit power at the \glspl{ue} $\rho_{\rmUE}$ for the \gls{nlos} scenario depicted in Fig.~\ref{fig:2}(a), with \gls{ula} at the \gls{bs}.} \label{fig:NMSE_vs_SNR} \vspace{-3mm}
\end{figure}

Let us define the \gls{nmse} of the channel estimation for \gls{ue}~$k$ as
\begin{align}
\overline{\mathrm{NMSE}}_k \triangleq \Exp \left[\frac{\|\hat{\bar{\g}}_{k} - \bar{\g}_{k}\|^2}{\|\bar{\g}_{k}\|^2}\right]
\end{align}
for the effective channels $\{\hat{\bar{\g}}_k\}_{k=1}^{K}$ resulting from \name{} and as
\begin{align}
\mathrm{NMSE}_k \triangleq \frac{1}{N} \sum_{n=1}^N \Exp \left[\frac{\|\hat{\g}_{k,n} - \g_{k,n}\|^2}{\|\g_{k,n}\|^2}\right]
\end{align}
for the channel matrices $\{\hat{\H}_k\}_{k=1}^{K}$. Considering the \gls{nlos} scenario in Fig.~\ref{fig:2}(a), Fig.~\ref{fig:NMSE_vs_SNR} plots the \gls{nmse} of the channel estimation versus the transmit power at the \glspl{ue}.\footnote{Note that the transmit \gls{snr} is given by simply dividing the transmit power at the \gls{bs} by the noise variance at the \gls{bs}, which is given in Table~\ref{tab:param}.} Thanks to the improved statistical orthogonality between the \glspl{ue}, \name{} considerably increases the channel estimation accuracy with respect to the case where the original \gls{mimo} channels are estimated in the presence of pilot contamination among the antennas of different \glspl{ue}.

\begin{figure}[t]
\centering
\includegraphics[scale=0.8]{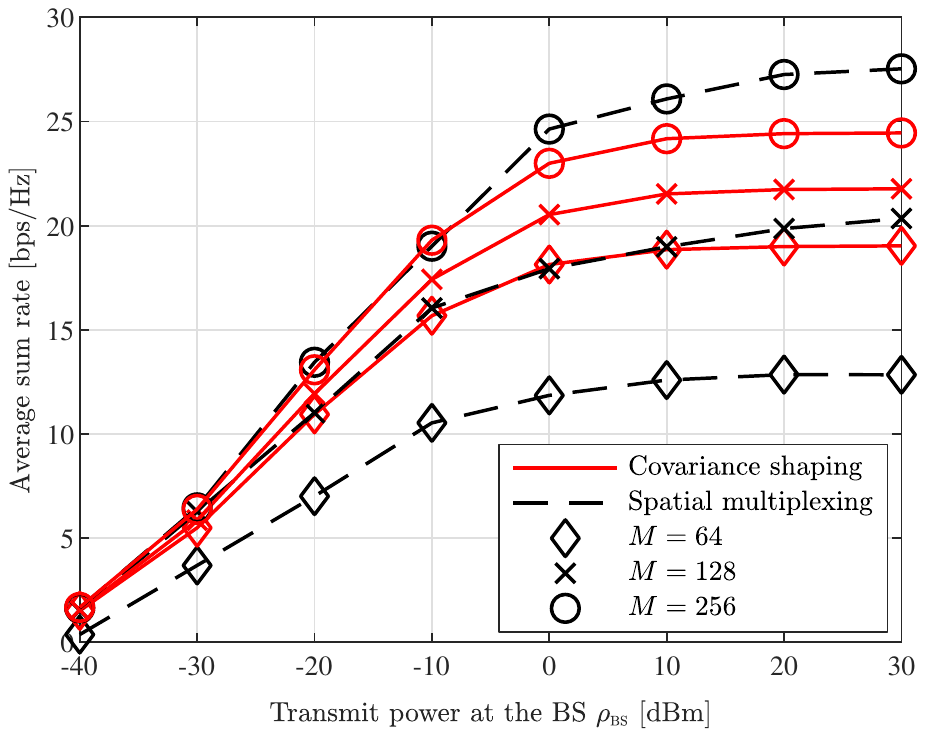}
\caption{Two-\gls{ue} case: average sum rate versus the transmit power at the \gls{bs} $\rho_{\rmBS}$ for the \gls{nlos} scenario depicted in Fig.~\ref{fig:2}(a), with $d=4$~m, \gls{ula} at the \gls{bs}, and for different values of $M$.} \label{fig:srate_SNR_2UE}
\end{figure}
\begin{figure}[t]
\centering
\includegraphics[scale=0.8]{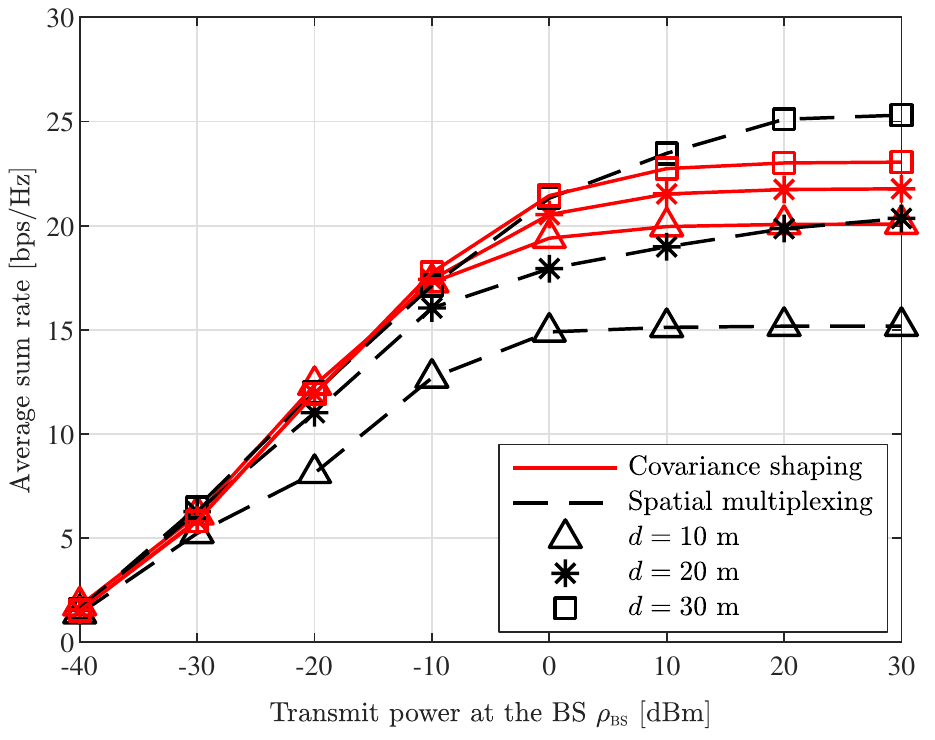}
\caption{Two-\gls{ue} case: average sum rate versus the transmit power at the \gls{bs} $\rho_{\rmBS}$ for the \gls{los} scenario depicted in Fig.~\ref{fig:3}(a), with $M=128$, \gls{ula} at the \gls{bs}, and for different values of $d$.} \label{fig:srate_corr} \vspace{-3mm}
\end{figure}

Considering again the \gls{nlos} scenario in Fig.~\ref{fig:2}(a), Fig.~\ref{fig:srate_SNR_2UE} plots the average sum rate versus the transmit power at the \gls{bs} $\rho_{\rmBS}$ for different numbers of \gls{bs} antennas $M$. For $M \leq 128$, it is straightforward to see that the signal subspace separation enforced by \name{} during both the uplink pilot-aided channel estimation phase and the downlink data transmission phase has a highly beneficial effect on the system performance. However, \name{} is outperformed by the reference scheme in~(B) for $M = 256$. In this setting, the enhanced spatial selectivity of the \gls{bs} allows to effectively separate the \glspl{ue} and promotes the transmission of multiple data streams per \gls{ue}. Focusing now on the \gls{los} scenario in Fig.~\ref{fig:3}(a), Fig.~\ref{fig:srate_corr} plots the average sum rate versus the transmit power at the \gls{bs} $\rho_{\rmBS}$, with $M=128$ and for different values of the inter-\gls{ue} distance $d$. Evidently, the reference scheme in~(B) is markedly limited by the overlapping channel covariance matrices of the \glspl{ue} and is able to effectively separate the \glspl{ue} only for very large inter-\gls{ue} distances (i.e., $d=30$~m). On the other hand, \name{} is able to enforce statistical orthogonality even when the \glspl{ue} are closely spaced.

\subsection{Multi-UE Case} \label{subsec:NR_multiUE}

\begin{figure}[t]
\centering
\begin{subfigure}{0.45\textwidth}
\centering
\vspace{0.3mm}
\includegraphics[scale=0.59]{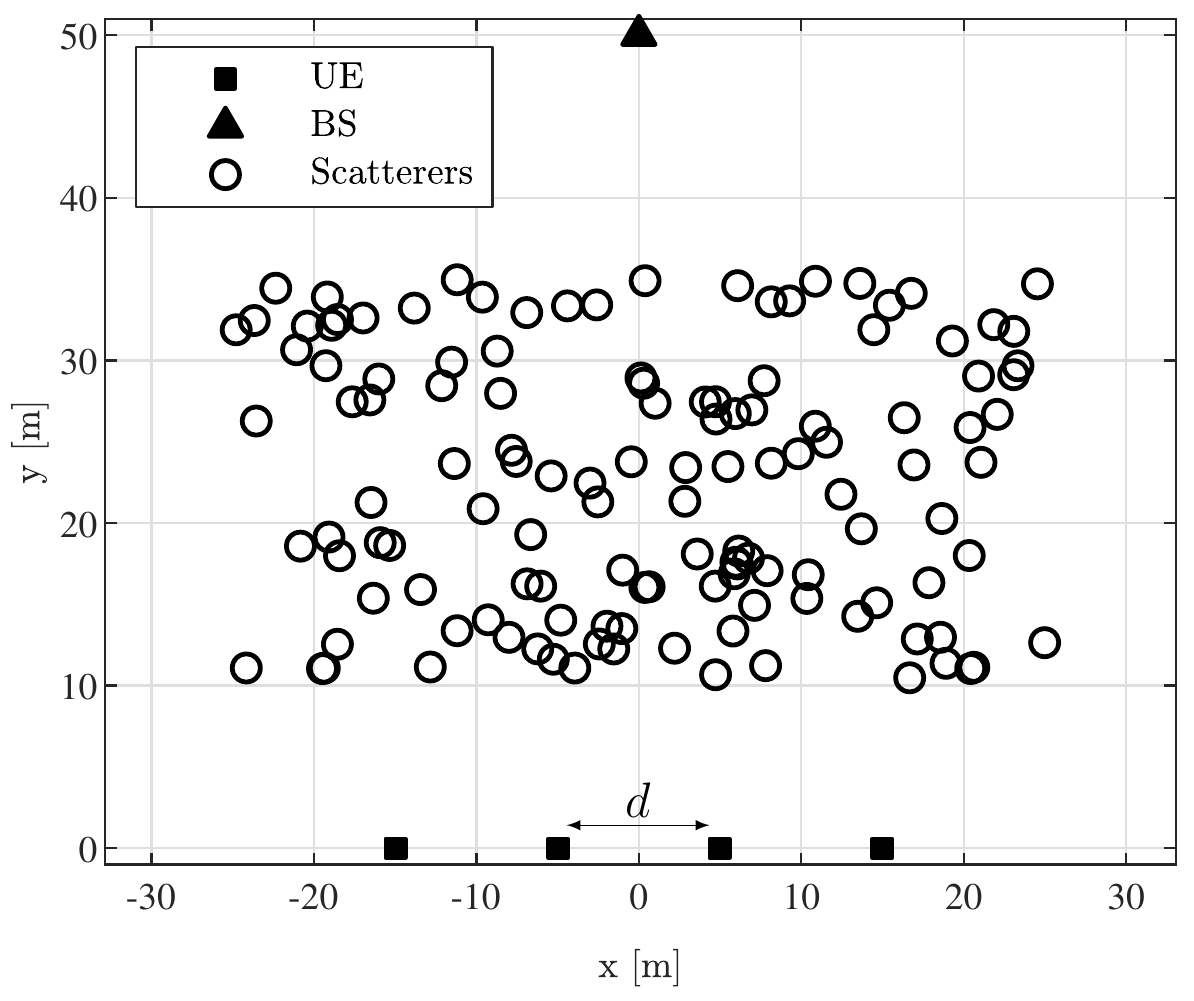}
\caption{NLoS scenario with $K=4$.}
\label{fig:map_nlos4}
\end{subfigure} \hfill
\begin{subfigure}{0.45\textwidth}
\centering
\vspace{7.7mm}
\hspace{4.4mm} \includegraphics[scale=0.85]{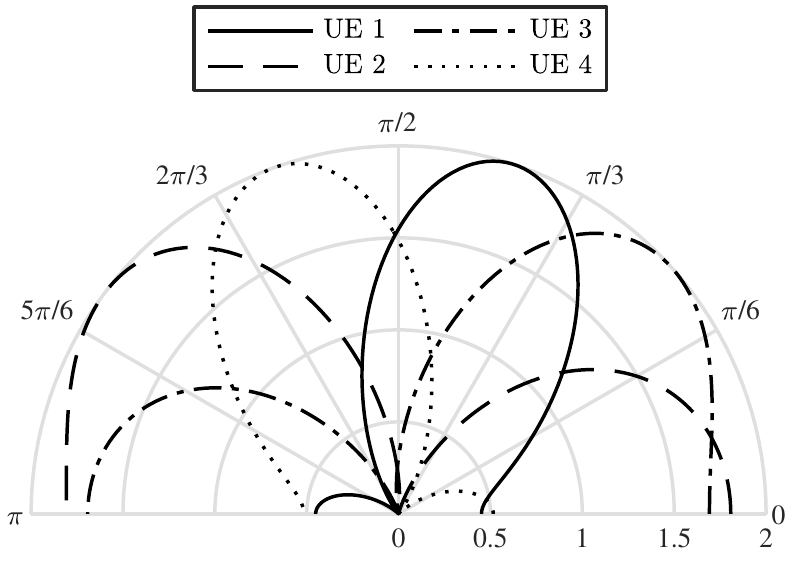}
\caption{Corresponding \name{} vectors.}
\label{fig:v_vectors}
\end{subfigure}
\caption{Multi-\gls{ue} case: (a) 2D map of the considered \gls{nlos} scenario with $K=4$ and variable inter-\gls{ue} distance $d$; (b) corresponding \name{} vectors with $N=2$ obtained with Algorithm~\ref{alg:A2}.} \label{fig:6} \vspace{-3mm}
\end{figure}

\begin{figure}[t]
\centering
\includegraphics[scale=0.8]{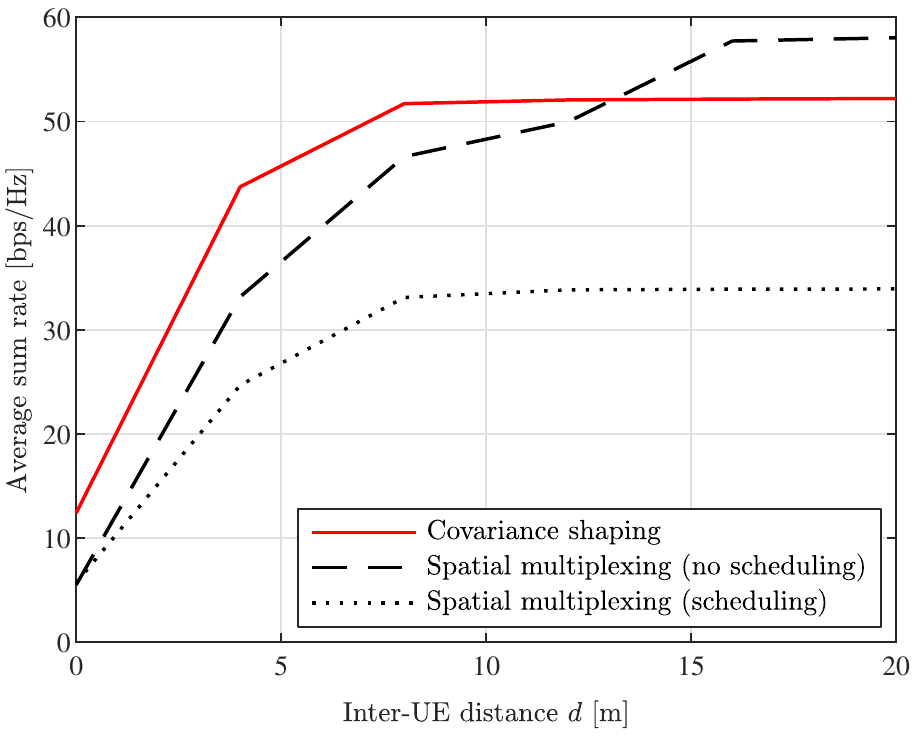}
\caption{Multi-\gls{ue} case: average sum rate versus the inter-\gls{ue} distance $d$ for the \gls{nlos} scenario depicted in Fig.~\ref{fig:6}(a), with $M=128$, \gls{ula} at the \gls{bs}, and $\rho_{\rmBS}=30$~dBm.} \label{fig:srate_K} \vspace{-3mm}
\end{figure}

Following the discussion in Section~\ref{subsec:MultiUE}, we now examine the general case of $K \geq 2$. We begin by considering the \gls{nlos} scenario in Fig.~\ref{fig:6}(a) with $K=4$. Here, the \gls{bs} is equipped with a \gls{ula} and we vary the inter-\gls{ue} distance $d$ in order to quantify its impact on the degree of statistical orthogonality among the \glspl{ue}. In this setting, $P=2$ orthogonal pilots are assigned such that two adjacent \glspl{ue} always utilize orthogonal pilots. The \name{} vectors obtained with Algorithm~\ref{alg:A2} are shown in Fig.~\ref{fig:6}(b): these tend to focus their power along reflected paths that are as orthogonal as possible to each other while also carrying sufficient channel power (cf. Fig.~\ref{fig:2}(b)). Fig.~\ref{fig:srate_K} plots the average sum rate versus the inter-\gls{ue} distance $d$, with $M=128$ and $\rho_{\rmBS}=30$~dBm. Note that, under the considered pilot assignment, pilot contamination occurs only between \glspl{ue} at distance $2 d$. In this context, we also study the case where the \glspl{ue} are scheduled into two separate groups of non-adjacent \glspl{ue} without pilot contamination (scheduling), as opposed to the case where all the \glspl{ue} are the served simultaneously with pilot contamination (no scheduling). As in Fig.~\ref{fig:srate_corr}, the reference scheme in~(B) outperforms \name{} only for very large inter-\gls{ue} distances (i.e., $d \geq 13$~m). In this regard, the scheduling approach is shown to further deteriorate the performance of spatial multiplexing due to the pre-log factor of $1/2$ in the sum rate despite avoiding the pilot contamination.

\begin{figure}[t]
\centering
\includegraphics[scale=0.8]{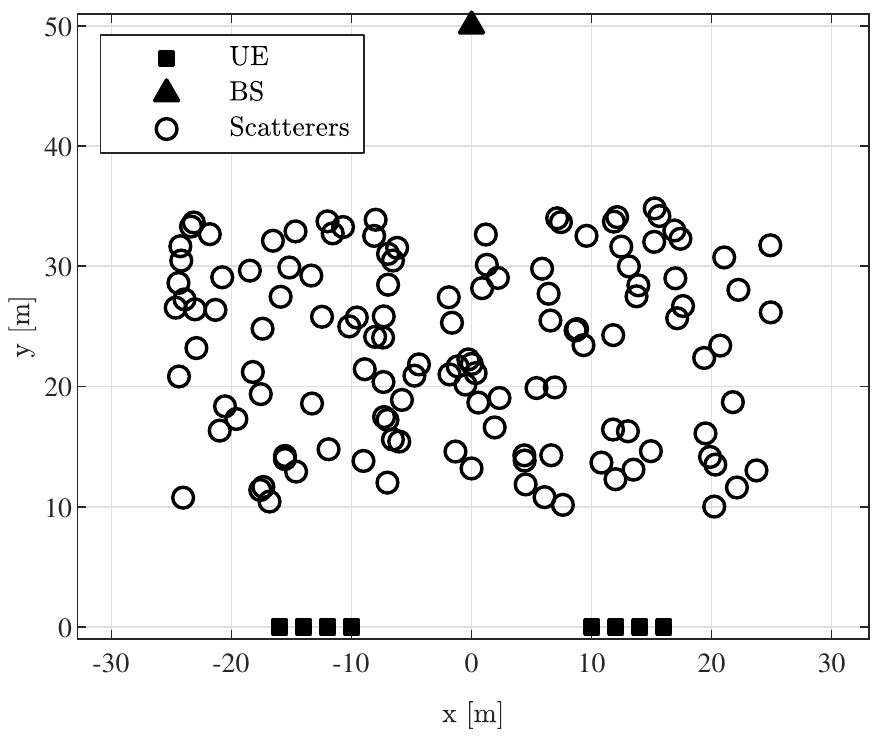}
\caption{Multi-\gls{ue} case: (a) 2D map of the considered \gls{nlos} scenario with $K=8$, inter-\gls{ue} distance $d=2$~m, and inter-group distance $D=20$~m.} \label{fig:8}
\end{figure}

\begin{figure}[t]
\centering
\includegraphics[scale=0.8]{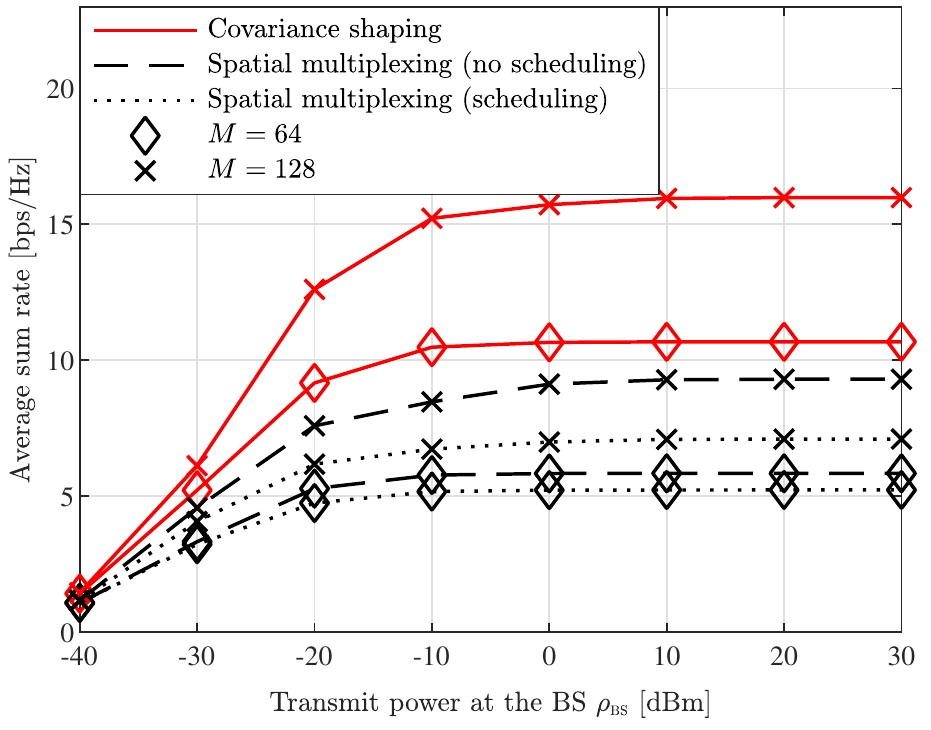}
\caption{Multi-\gls{ue} case: average sum rate versus the transmit power at the \gls{bs} $\rho_{\rmBS}$ for the \gls{nlos} scenario depicted in Fig.~\ref{fig:8}, with \gls{upa} at the \gls{bs} and for different values of $M$.} \label{fig:srate_SNR} \vspace{-3mm}
\end{figure}

We conclude this section by considering the \gls{nlos} scenario in Fig.~\ref{fig:8} with $K=8$. Here, the \gls{bs} is equipped with a \gls{upa} and is placed at a height of $20$~m with respect to the \glspl{ue} and the scatterers. Moreover, the \glspl{ue} are separated into two groups with inter-group distance $D=20$~m and the inter-\gls{ue} distance is $d=2$~m, which results in highly overlapping channel covariance matrices within the same group. In this setting, $P=4$ orthogonal pilots are assigned such that two adjacent \glspl{ue} always utilize orthogonal pilots and \name{} is applied independently to each group. Fig.~\ref{fig:srate_SNR} plots the average sum rate versus the transmit power at the \gls{bs} $\rho_{\rmBS}$ for different numbers of \gls{bs} antennas $M$. Note that $M=64$ corresponds to $M_x=M_y=8$, whereas $M=128$ corresponds to $M_x=16$ and $M_y=8$. As in Fig.~\ref{fig:srate_K}, we also study the case where the \glspl{ue} in each group are scheduled into two separate subgroups of non-adjacent \glspl{ue} without pilot contamination among groups (scheduling). The signal subspace separation enforced by \name{} during both the uplink pilot-aided channel estimation phase and the downlink data transmission phase becomes even more beneficial when the \gls{bs} is equipped with a \gls{upa} with limited spatial selectivity in the azimuth direction. Again, the scheduling approach is shown to further deteriorate the performance of spatial multiplexing due to the pre-log factor of $1/2$ in the sum rate despite avoiding the pilot contamination within the same group.

\section{Conclusions} \label{sec:Con}

In this paper, we introduce the novel concept of \gls{mimo} \name{} as a means to achieve statistical orthogonality among interfering \glspl{ue}. It consists in preemptively applying a statistical beamforming at each \gls{ue} during both the uplink pilot-aided channel estimation phase and the downlink data transmission phase, aiming at enforcing a separation of the signal subspaces of the \glspl{ue} that would be otherwise highly overlapping. The proposed \gls{mimo} \name{} framework exploits the realistic non-Kronecker structure of massive \gls{mimo} channels, which allows to suitably alter the channel statistics perceived at the \gls{bs} by designing the transceiver at the \gls{ue}-side. To compute the \name{} strategies, we present a low-complexity block coordinate descent algorithm that minimizes the inter-\gls{ue} interference (as a metric to measure the spatial correlation), which is proved to converge to a limit point of the original nonconvex problem or to a stationary point in the two-UE case. We provide numerical results characterizing several scenarios where \gls{mimo} \name{} outperforms a reference schemes employing the multiple antennas at the \gls{ue} for spatial multiplexing. Specifically, this occurs when the spatial selectivity of the \gls{bs} is not sufficient to separate \glspl{ue} exhibiting high spatial correlation, (e.g., when the \gls{bs} is equipped with a \gls{upa} with limited spatial selectivity in the azimuth direction) and the channel estimation is limited by strong pilot contamination. Future work will consider \gls{mimo} \name{} in combination with multi-stream transmission and will analyze the trade-off between statistical orthogonality and spatial multiplexing.

\begin{appendices}

\section{Derivations of the Effective SINR in \texorpdfstring{\eqref{eq:sinr_ce}}{}} \label{sec:app2}

In this Appendix, we derive the expectation terms in \eqref{eq:sinr_ce_0} to obtain the expression of the effective \gls{sinr} in \eqref{eq:sinr_ce}. First of all, given the multi-\gls{ue} effective channel matrix $\bar{\H} = [\bar{\g}_1^{\tran}, \ldots, \bar{\g}_K^{\tran}]^{\tran}$, where the effective channels $\{ \bar{\g}_k \}_{k=1}^{K}$ are independent and $\bar{\g}_k \sim \setC\setN(\0,\bar{\Phib}_k)$, it is easy to verify that
\begin{align}
\Exp[\|\bar{\H}\|_{\rmm{F}}^2] & = \sum_{k=1}^{K} \tr (\bar{\Phib}_k).
\end{align}
Furthermore, recalling the definition of $\hat{\bar{\g}}_{k}$ in \eqref{eq:g_hat}, we have
\begin{align} \label{eq:der1}
\Exp[\bar{\g}_k \hat{\bar{\g}}_k^{\herm}] = \tr (\bar{\Phib}_k \Q_k^{-1} \bar{\Phib}_k).
\end{align}
We now focus on deriving $\Var[\bar{\g}_k\hat{\bar{\g}}_k^{\herm}]$, with $k \in \setS_{p}$. We have
\begin{align}
& \Var[\bar{\g}_k\hat{\bar{\g}}_k^{\herm}] \nonumber \\
& = \Exp \big[ |\bar{\g}_k\hat{\bar{\g}}_k^{\herm}|^2 \big] - \big|\Exp[\bar{\g}_k\hat{\bar{\g}}_k^{\herm}]\big|^2 \\
& = \Exp\bigg[\bar{\g}_k\bar{\Phib}_k\Q_k^{-1}\bigg(\bar{\g}_k^{\herm}+\sum_{j\in \setS_p\setminus\{k\}} \bar{\g}_j^{\herm}+\frac{1}{\tau \sqrt{\rho_{\rmUE}}} \Z \p_p^{\herm} \bigg) \nonumber \\
& \phantom{=} \ \times \bigg(\bar{\g}_k+\sum_{j\in \setS_p\setminus\{k\}} \bar{\g}_j+\frac{1}{\tau \sqrt{\rho_{\rmUE}}} \p_p \Z^{\herm} \bigg) \Q_k^{-1}\bar{\Phib}_k\bar{\g}_k^{\herm}\bigg] \nonumber \\
& \phantom{=} \ - \tr (\bar{\Phib}_k \Q_k^{-1} \bar{\Phib}_k)^{2} \label{eq:der2} \\
& =\Exp[\bar{\g}_k\bar{\Phib}_k\Q_k^{-1}\bar{\g}_k^{\herm}\bar{\g}_k\Q_k^{-1}\bar{\Phib}_k\bar{\g}_k^{\herm}]\nonumber \\
&\phantom{=} \ +\sum_{j\in \setS_p\setminus\{k\}} \Exp[\bar{\g}_k\bar{\Phib}_k\Q_k^{-1}\bar{\g}_j^{\herm}\bar{\g}_j\Q_k^{-1}\bar{\Phib}_k\bar{\g}_k^{\herm}] \nonumber \\
& \phantom{=} \ + \frac{1}{\tau^{2} \rho_{\rmUE}} \Exp[\bar{\g}_k\bar{\Phib}_k\Q_k^{-1} \Z \p_p^{\herm} \p_p \Z^{\herm} \Q_k^{-1}\bar{\Phib}_k\bar{\g}_k^{\herm}] \nonumber \\
& \phantom{=} \ - \tr (\bar{\Phib}_k \Q_k^{-1} \bar{\Phib}_k)^{2} \label{eq:der3}
\end{align}
where in \eqref{eq:der2} we have used the expression in \eqref{eq:der1} and where \eqref{eq:der3} follows from the independence between $\bar{\g}_k$ and $\bar{\g}_j$, $\forall k \neq j$, and between $\bar{\g}_k$ and $\Z$. Then, we write $\bar{\g}_k^{\herm} = \bar{\Phib}_k^{\frac{1}{2}} \x_{k}$, with $\x_{k} \sim \setC \setN (0,\I_M)$, and obtain
\begin{align}
& \Var[\bar{\g}_k\hat{\bar{\g}}_k^{\herm}] \nonumber \\
& = \Exp \big[ \x_{k}^{\herm} \bar{\Phib}_k^{\frac{1}{2}}\bar{\Phib}_k\Q_k^{-1}\bar{\Phib}_k^{\frac{1}{2}}\x_{k} \x_{k}^{\herm} \bar{\Phib}_k^{\frac{1}{2}}\Q_k^{-1}\bar{\Phib}_k\bar{\Phib}_k^{\frac{1}{2}}\x \big] \nonumber \\
& \phantom{=} \ + \sum_{j\in \setS_p\setminus\{k\}} \tr (\bar{\Phib}_k^2\Q_k^{-1}\bar{\Phib}_j\Q_k^{-1}\bar{\Phib}_k) \nonumber \\
& \phantom{=} \ + \frac{1}{\tau \varrho_{\rmUE}} \tr (\bar{\Phib}_k^2\Q_k^{-2}\bar{\Phib}_k) - \tr (\bar{\Phib}_k \Q_k^{-1} \bar{\Phib}_k)^{2} \label{eq:der4} \\
& = \tr (\bar{\Phib}_k^2\Q_k^{-1}\bar{\Phib}_k\Q_k^{-1}\bar{\Phib}_k) + \tr (\bar{\Phib}_k\Q_k^{-1}\bar{\Phib}_k)^2 \nonumber \\
& \phantom{=} \ + \sum_{j\in \setS_p\setminus\{k\}} \tr (\bar{\Phib}_k^2\Q_k^{-1}\bar{\Phib}_j\Q_k^{-1}\bar{\Phib}_k) \nonumber \\
& \phantom{=} \ + \frac{1}{\tau \varrho_{\rmUE}} \tr (\bar{\Phib}_k^2\Q_k^{-2}\bar{\Phib}_k) - \tr (\bar{\Phib}_k\Q_k^{-1}\bar{\Phib}_k)^2 \label{eq:der5} \\
& = \tr\bigg(\bar{\Phib}_k^2\Q_k^{-1} \bigg( \bar{\Phib}_k+\sum_{j\in \setS_p\setminus\{k\}} \bar{\Phib}_j+ \frac{1}{\tau \varrho_{\rmUE}}\I_M\bigg) \Q_k^{-1}\bar{\Phib}_k\bigg) \\
& = \tr (\bar{\Phib}_k^2\Q_k^{-1}\bar{\Phib}_k) \label{eq:der6}
\end{align}
where \eqref{eq:der4} is obtained by substituting $\Exp[\Z \p_p^{\herm} \p_p \Z^{\herm}] = \tau \sigma_{\rmBS}^2 \I_M$, in \eqref{eq:der5} we have exploited the fact that $\Exp[\x_{k} \x_{k}^{\herm}\A \x_{k} \x_{k}^{\herm}] = \A + \tr (\A) \I_M$ for a given Hermitian matrix $\A \in \Compl^{M\times M}$ \cite[App.~A.2]{Mar16}, and in \eqref{eq:der6} we have used the definition of $\Q_{k}$ provided in Section~\ref{subsec:CS_UL_CE}. Finally, we focus on deriving $\Exp \big[ |\bar{\g}_k\hat{\bar{\g}}_j^{\herm}|^2 \big]$, with $k \in \setS_{p}$ and with $j \in \setS_{q}$. Following similar steps as above, we have
\begin{align}
& \Exp \big[ |\bar{\g}_k\hat{\bar{\g}}_j^{\herm}|^2 \big] \nonumber \\
& = \Exp \bigg[ \bar{\g}_k \bar{\Phib}_j \Q_j^{-1} \bigg( \bar{\g}_j^{\herm} + \sum_{l \in \setS_q \setminus \{j\}} \bar{\g}_l^{\herm} + \frac{1}{\tau \sqrt{\rho_{\rmUE}}} \Z \p_q^{\herm} \bigg) \nonumber \\
& \phantom{=} \ \times \bigg( \bar{\g}_j + \sum_{l \in \setS_q\setminus\{j\}} \bar{\g}_l + \frac{1}{\tau \sqrt{\rho_{\rmUE}}} \p_q \Z^{\herm} \bigg) \Q_j^{-1} \bar{\Phib}_j \bar{\g}_k^{\herm} \bigg] \\
& = \Exp[\bar{\g}_k \bar{\Phib}_j \Q_j^{-1} \bar{\g}_j^{\herm} \bar{\g}_j \Q_j^{-1} \bar{\Phib}_j \bar{\g}_k^{\herm}] \nonumber \\
& \phantom{=} \ +\sum_{l \in \setS_q \setminus \{j\}} \Exp[\bar{\g}_k \bar{\Phib}_j \Q_j^{-1} \bar{\g}_l^{\herm} \bar{\g}_l \Q_j^{-1} \bar{\Phib}_j \bar{\g}_k^{\herm}] \nonumber \\
& \phantom{=} \ + \frac{1}{\tau^{2} \rho_{\rmUE}} \Exp[\bar{\g}_k\bar{\Phib}_k\Q_k^{-1} \Z \p_q^{\herm} \p_q \Z^{\herm} \Q_k^{-1}\bar{\Phib}_k\bar{\g}_k^{\herm}] \\
& = \tr(\bar{\Phib}_k \bar{\Phib}_j \Q_j^{-1} \bar{\Phib}_j \Q_j^{-1} \bar{\Phib}_j) \nonumber \\
& \phantom{=} \ + \sum_{l \in \setS_q \setminus \{j\}} \tr(\bar{\Phib}_k \bar{\Phib}_j \Q_j^{-1} \bar{\Phib}_l \Q_j^{-1} \bar{\Phib}_j) \nonumber \\
& \phantom{=} \ + \mathbbm{1}_{q=p} \tr(\bar{\Phib}_k \Q_j^{-1} \bar{\Phib}_j)^{2} + \frac{1}{\tau \varrho_{\rmUE}} \tr(\bar{\Phib}_k \bar{\Phib}_j \Q_j^{-2} \bar{\Phib}_j) \\
& = \tr \bigg( \bar{\Phib}_k \bar{\Phib}_j \Q_j^{-1} \bigg( \bar{\Phib}_j + \sum_{l \in \setS_q \setminus \{j\}} \bar{\Phib}_l + \frac{1}{\tau \varrho_{\rmUE}} \I_{M} \bigg) \Q_j^{-1} \bar{\Phib}_j \bigg) \nonumber \\
& \phantom{=} \  + \mathbbm{1}_{q=p} \tr(\bar{\Phib}_k \Q_j^{-1} \bar{\Phib}_j)^{2} \\
& = \tr (\bar{\Phib}_k \bar{\Phib}_j \Q_j^{-1} \bar{\Phib}_j) + \mathbbm{1}_{q=p} \tr(\bar{\Phib}_k \Q_j^{-1} \bar{\Phib}_j)^{2}. \label{eq:der7}
\end{align}
Note that the second term in \eqref{eq:der7} is active only if \glspl{ue}~$k$ and~$j$ are assigned the same pilot.

\end{appendices}

\addcontentsline{toc}{chapter}{References}
\bibliographystyle{IEEEtran}
\bibliography{IEEEabrv,refs}

\end{document}